\documentclass{aa} 

\makeatletter
\renewcommand\thelinenumber{}
\renewcommand\makeLineNumber{}
\makeatother

\usepackage{graphicx}

\usepackage{newtxtext,newtxmath}

\usepackage{subcaption}
\usepackage{placeins}

\usepackage{svg}
\usepackage{xcolor}
\usepackage{bm}

\usepackage{booktabs}
\usepackage[normalem]{ulem}

\svgpath{{./plots/}}

\usepackage{siunitx}
\usepackage[colorlinks=true, linkcolor=blue, citecolor=blue, urlcolor=blue]{hyperref}

\begin{document}

   \title{Emulating redshift mixing due to blending \\ in weak gravitational lensing}



   \author{
        Zekang Zhang \inst{1} \thanks{\email{zekang.zhang@physik.lmu.de}} \and
        Daniel Gruen \inst{1,2} \and 
        Luca Tortorelli \inst{1} \and 
        Shun-Sheng Li \inst{3,4} \and
        Jamie McCullough \inst{5}
        }

   \institute{
        Universitäts-Sternwarte, Fakultät für Physik, Ludwig-Maximilians Universität München, Scheinerstr. 1, 81679 München, Germany \\ 
        \and Excellence Cluster ORIGINS, Boltzmannstr. 2, 85748 Garching, Germany \\ 
        \and Ruhr University Bochum, Faculty of Physics and Astronomy, Astronomical Institute (AIRUB), German Centre for Cosmological Lensing, 44780 Bochum, Germany\\
        \and Leiden Observatory, Leiden University, Einsteinweg 55, 2333 CC Leiden, the Netherlands\\
        \and Department of Astrophysical Sciences, Princeton University, Princeton, NJ 08544, USA\\ }

   \date{Received XXX}

 
  \abstract
   {Galaxies whose images overlap in the focal plane of a telescope, commonly referred to as blends, are often located at different redshifts. Blending introduces a challenge to weak-lensing cosmology probes since such blends are subject to shear signals from multiple redshifts.}
   {This effect can be described by joining shear bias and redshift characterisation in the effective redshift distribution, $n_{\gamma}(z)$, which includes the response of apparent shapes of detected objects to shear of galaxies at redshift, $z$. In this work, we propose a novel method to correct $n_{\gamma}(z)$ for redshift-mixed blending by emulating the shear response to neighbouring galaxies.} 
   {We designed a `half-sky-shearing' simulation with Subaru Hyper Suprime Cam (HSC) wide-like specifications,  which allowed us to extract the response of a detected object's measured ellipticity to the shearing of neighbouring galaxies among numerous galaxy pairs.}
   {We demonstrate the feasibility of accurately emulating these pairwise responses and validate the robustness of our approach under varying observing conditions and galaxy population uncertainties. We find that the effective redshift of sources at the high-redshift tail of the distribution is about 0.05 lower than expected when the effect is not modelled.}
   {Given adequately processed image simulations, our correction method can be readily incorporated into future cosmological analyses to mitigate this source of systematic error.}

   \keywords{Gravitational lensing: weak --
                large-scale structure of Universe --
                Methods: numerical
               }

   \maketitle



\section{Introduction}

The gravitational field of the cosmic large-scale structure disturbs the observed shapes of background galaxies, typically at the per cent level, in a phenomenon known as cosmic shear. The coherent shapes of galaxy ensembles over a wide sky area contain valuable information about fluctuations of the underlying matter density and their evolution \citep{weinbergObservationalProbesCosmic2013,kilbingerCosmologyCosmicShear2014}. Weak lensing has become one of the most important probes for testing cosmological models and is a key focus of various wide-field surveys. Major Stage-III experiments include the Dark Energy Survey (DES; e.g. \citealt{AbbottDESY3Results2022}),  Kilo-Degree Survey (KiDS; e.g. \citealt{HeymansKiDS10002021, WrightKiDSLegacy2025}), and Subaru Hyper Suprime-Cam (HSC; e.g. \citealt{MoreHSCY3Results2023}). Several new analyses such as Euclid \citep{LaureijsEuclidReport2011,MellierEuclid2024}, Chinese Space Station Telescope (CSST; \citealt{ZhanCSSTCosmology2019}),  Nancy
Grace Roman Space Telescope \citep{SpergelRom2015}, and Legacy Survey of Space and Time (LSST; \citealt{IvezicLSST2019}) are underway. With the ever-increasing survey volume, the control of systematics becomes the leading challenge. As a result, the development and testing of accurate methods for estimating and interpreting cosmic shear signals has  become an active field of research \citep{mandelbaumWeakLensingPrecision2018}.

Shear estimation is subject to a range of systematic uncertainties, with image noise (e.g. \citealt{melchiorMeansConfusionHow2012, refregierNoiseBiasWeak2012}) and point spread function (PSF) modelling errors (e.g. \citealt{paulin-henrikssonPSFCalibrationRequirements2008}) typically being the dominant sources. Other issues, such as an inaccurate modelling of galaxy morphology (e.g. \citealt{voigtLimitationsModelfittingMethods2010}) and shear-dependent selection effects (e.g. \citealt{fenechcontiCalibrationWeaklensingShear2017,sheldonPracticalWeaklensingShear2017}), have received increasing attention. Innovative algorithms have been developed to address all or most of the above issues (e.g. \citealt{bernsteinBayesianLensingShear2014, ZhangShear2015, LiFPFS2023}). Among them, \textsc{Metacalibration} \citep{huffMetacalibration2017,sheldonPracticalWeaklensingShear2017}, which directly calibrates measured shear on individual galaxy observations, can, in principle, correct most of pixel-level biases.

Beyond these challenges, blending, where two or more sources overlap on the image plane, has become increasingly significant as surveys reach greater depths. This issue is especially severe for ground-based observations due to atmospheric seeing. DES reported that approximately 30\% of galaxies used for shear measurements are affected by light from neighbouring sources \citep{DESOverview2016}. This fraction increases dramatically to about 60\% in deeper surveys such as HSC \citep{BoschHSC2018} and LSST \citep{SanchezOverlapping2021,MelchiorBlending2021}.

The mixing of light from multiple galaxies in blending complicates galaxy identification and photometric measurements, impacting downstream analyses such as photometric redshift estimation and shear inference. Most intuitively, the light contamination from neighbouring galaxies alters a galaxy's apparent shape \citep{samuroffDarkEnergySurvey2018}. Even faint sources below the detection threshold are ubiquitous and have been reported to have an impact on shape measurement \citep{hoekstraCanadianClusterComparison2015,hoekstraStudySensitivityShape2017,EuclidBlending2019}; for example, by influencing the determination of the background. Besides shape measurement, \cite{sheldonMitigatingSheardependentObject2020} demonstrated that shear could alter the post-seeing topology of blended galaxies; namely, that it could depend on their alignment relative to the shear direction. This effect makes object detection shear-dependent and has been reported to contribute to a non-negligible multiplicative shear bias. While state-of-the-art de-blending algorithms are capable of separating most identified blended sources \citep{DeblendingDenclue2016, drlica-wagnerDarkEnergySurvey2018, MelchiorScarlet2018}, ambiguous cases (where only a single peak is detected in a blended group) pose an irreducible challenge. LSST anticipates that 10–30\% of its detections will be ambiguous blends \citep{dawsonEllipticityDistributionAmbiguously2014,TroxelJoint2023}. They are shown to increase dispersion in galaxy ellipticities and alter their distribution \citep{dawsonEllipticityDistributionAmbiguously2014}, which some calibration approaches have used as a prior. \cite{NourbakhshBlendingCosmology2022} found that these blends can bias the derived value of $S_8$ by up to $2\sigma$ in tomographic analyses using mock catalogues, although they have a relatively minor impact on photometric redshifts. \cite{GencBlendDetect2025} found a 2\% underestimation of $S_8$ due to detection bias correlating with matter as ambiguous blends occur preferably with high foreground density. \cite{LiangBlend2025} showed that catalogue-level information can be used to remove 30\% (80\%) of ambiguous blends, albeit at the cost of discarding 10\% (50\%) of the entire sample.

In principle, these blending-induced biases can all be studied using image simulations. The overall shear bias is usually captured by shearing the input galaxies with a constant amount (e.g. \citealt{fenechcontiCalibrationWeaklensingShear2017,MandelbaumImSim2018,KannawadiImSim2019,KacprzakImSim2020}). This necessitates a realistic and comprehensive reconstruction of observational effects and galaxy properties, including their morphology and abundance as a function of redshift. Developed on the basis of \textsc{Metacalibration}, we can now use \textsc{Metadetection} \citep{sheldonMitigatingSheardependentObject2020}, which introduces artificial shear to a patch of the observed sky prior to detection. It has been demonstrated to effectively minimise the shear-dependent detection bias directly from data (see also \citealt{HoekstraDetection2021}). However, since these methods introduce the same shear to all galaxies, they ignore the complexity when components within a blended system are located at different redshifts \citep{zuntzDarkEnergySurvey2018,maccrannY3ResultsBlending2022, liKiDSLegacyCalibrationUnifying2023}. The shear they experience differs due to the projected matter fields and lensing kernels. Light contamination from a neighbouring galaxy at a different redshift alters the measured signal.

This raises a distinct and more recently recognised challenge, which has been addressed in the latest efforts using image simulations with redshift-dependent shear signals. \cite{maccrannY3ResultsBlending2022} divided simulated galaxies into four redshift bins and applied shear to each bin individually. The shear responses of detections associated with galaxies in other bins are measured and fit with a redshift-dependent model to correct the \textsc{Metacalibration}-response weighted galaxy redshift distribution. Instead, \cite{liKiDSLegacyCalibrationUnifying2023} simulated multi-band images with a realistic, redshift-dependent variable shear field to jointly calibrate shear and redshift biases, aiming to account for these blending-induced effects. By comparing results from variable-shear simulations with constant-shear ones, they demonstrated clear redshift-dependent blending-induced shear biases. Several caveats apply to these simulation-based calibrations: a simulation in fine redshift slices is needed to achieve better fitting and becomes extremely costly, as the number of simulation sets increases proportionally with the redshift steps. Instead, incorporating a realistic shear field renders the calibration dependent on cosmology. Accounting for galaxy population uncertainty is also challenging, especially at the faint end of the luminosity function as we reach greater depth.

In this work, we introduce a new method to forward‐model redshift crosstalk from blending at the catalogue level. This allows us to compute a lensing‐weighted redshift distribution that captures signal leakage between redshift bins. Instead of measuring shear response between coarse redshift slices, we broke it down into responses between individual galaxy pairs, quantifying how much any neighbour contributes to an image's observed shape. Implementing this pair‐based model provides the flexibility to apply it to a full galaxy population by combining the expected pairs, to perform arbitrarily fine redshift calculations, and to consistently account for uncertainty in cosmological parameters and galaxy population, which is vital for accurate redshift correction yet prohibitively difficult with simulation‐based calibrations.

To construct such pairs, we simulated realistic images using the SKiLLS image simulator developed for KiDS \citep{liKiDSLegacyCalibrationUnifying2023} and sheared a random half of the input galaxies on the simulated sky, each in a randomised direction. By taking galaxy pairs consisting of a non-sheared galaxy and one of its neighbouring galaxies, the response is quantified as the difference in the measured shape of the detection associated with the non-sheared galaxy when the shear on the neighbour is either applied or not. With sufficiently large samples, this strategy explores all possible combinations of blends, allowing us to learn how the response relates to the properties of each pair. Since blending commonly includes non-detected neighbours and biases the measurements and the downstream processes relying on the measurements, we used the true properties of galaxies, which further necessitates emulating the selection process, including object detection and bin assignment.

In Sect. 2, we begin by reviewing the definition of the effective redshift distribution and introducing our formalism for describing redshift-dependent blending. In Sect. 3, we outline the input galaxy catalogue and the simulation strategy, including the organisation of the training and validation datasets, as well as the shape measurement method. In Sect. 4, we present the blending responses measured from simulations and demonstrate how we emulated their dependence on the properties of blended galaxy pairs and how we emulated the source detection using information about the neighbour galaxy. Additionally, we provide the results of the emulators, including the predicted effective redshift distributions of tomographic bins. We present our conclusions in Sect. 6.

\section{Effective redshift distribution for lensing in the presence of blending}
\label{formula}

We begin by considering a simplified case in which all galaxies are isolated in the image. Every object detected and included in a weak-lensing source sample can be unambiguously associated with a single galaxy in this setup. The ellipticity of a galaxy's associated detection is an estimator of the reduced shear, whose measurement is subject to various sources of bias. The multiplicative component of the bias can be expressed as a response, $\bm{R}$, of the measured ellipticity to the shear, $\bm{g}$, experienced by the galaxy. While $\bm{g}$ is a spin-2 quantity with two components and $\bm{R}$ is a $2\times2$ matrix, the two off-diagonal elements of $\bm{R}$ are empirically found to be negligible, and the diagonal elements are approximately equal {\citep{sheldonPracticalWeaklensingShear2017}}. Therefore, we are only concerned with a single component and use scalar notations $g$ and $R$ unless otherwise specified, where $R$ depends not only on the apparent properties of the galaxy, but also on its local observing conditions, such as the PSF and background noise. We denote the collection of all properties that the response depends on as $\bm{G}$. 

Neglecting the galaxy shape noise and the correlation of the intrinsic shapes, the observed shear, $g_{\rm obs}$, of an ensemble of detections at angular direction, $\bm{\theta}$, averages over the true shear at this sky position and at redshift, $z$, weighted by the shear response and abundance of galaxies \citep{maccrannY3ResultsBlending2022}:
\begin{equation}
    g_{\rm obs}(\bm{\theta}) = \int dz~g(z,\bm{\theta})\int d\bm{G}~n_{\bm{G}}(z,\bm{\theta}) R(\bm{G}). \; 
\end{equation}
Here, $n_{\bm{G}}(z,\bm{\theta})$ presents the multi-dimensional distribution of the properties and redshift of galaxies associated with the detections included in the sample. In this formulation, contributions from higher order shear bias are conventionally neglected, an approximation that holds for cosmic shear where $g\ll1$ (e.g. \citealt{LiHighOrder2024,JansenHighorder2024}). The additive component of the shear bias is ignored as it is unrelated to blending. 

The dependence of $g_{\rm obs}$ on sky position $\bm{\theta}$ enables the use of common statistics such as $n$-point correlation functions. Additionally, $n_{\bm{G}}(z,\bm{\theta})$ can be specified for galaxies associated with those detections selected into a particular redshift bin, enabling a tomographic analysis. 

Although galaxies are assumed to be uniformly distributed across the sky, their observed quantities (including number density and shapes) vary spatially, due to inhomogeneous observing conditions, lensing magnification effect, and so on. While spatial variation systematics are an important consideration and are consistently handled by our method, we assume in the following that observing conditions and shear fields do not depend on sky direction and omit the dependence on $\bm{\theta}$ in the remainder of this work.

Marginalising over the properties of galaxies associated with detections assigned to a tomographic bin, their number distribution in true redshift is
\begin{equation}
    n(z) = \int d\bm{G}~n_{\bm{G}}(z). \; 
\end{equation}
Weighted by shear response, the effective redshift distribution in the isolated galaxy scenario can be defined as
\begin{equation}
    n_{\gamma}(z) = \int d\bm{G}~n_{\bm{G}}(z) R(\bm{G}).\;
    \label{eqn:ngamma}
\end{equation}

We note that in this work, we normalised $\int \mathrm{d}z \, n(z)=1,$ while there are conventions that take $\int \mathrm{d}z \,  n_{\gamma}(z)=1$. The normalisation of $n_{\gamma}(z)$ in this work is therefore equal to the mean response to a constant shear applied to the set of detections in the sample. This is equivalent in effect to normalising $n_{\gamma}$ to unity and modelling overall multiplicative bias separately when predicting a lensing signal. Our choice of normalisation, however, simplifies the form of the equations we derive in the following.

We  move on to the case where blending can occur. Here, it becomes necessary to distinguish a detection from a galaxy more carefully. The former refers to whatever we identify from the observed image and select into the sample of interest based on its measured features. In the isolated galaxy scenario considered above, a detection would be unambiguously associated with a single truly existing galaxy and its features would provide a measurement of some of the galaxy's properties independent of the properties of any other galaxies in the vicinity.  In the presence of blending, in contrast, the measurements made of a detected object are affected by the properties of more than one galaxy.\footnote{We ignore here the case of `false detections' that are not related to any sky object but rather, e.g. to an artefact or noise fluctuations. Data reduction, detection, and filtering choices generally should attempt to ensure that such detections are extremely rare in the catalogues used for science analyses.}

With such distinction being made, we further define a primary galaxy to be the unique sky object with a distinct redshift associated with each detection, and, where blending occurs, the secondary galaxies to be the additional, truly existing galaxies nearby the primary galaxy. In the data, we can possibly (but not necessarily) observe additional detections associated with the secondary galaxies, as there can be undetected blends very near the primary galaxy in projection or they might be so faint that they are not detected individually even when resolved. Their light affects the measured shape of the primary galaxy in a way that potentially changes when the secondaries are being sheared. Importantly, secondary galaxies are not necessarily placed at the same redshift as the primary galaxy, meaning they can experience a different shear. The mixture of light among different redshifts causes a cross-talk between shear signals in the measurement if the light cannot be perfectly separated.

Considering blending between different redshifts, the shear of secondary galaxies contributes an extra component in the measured signal,
\begin{align}
\label{eq:g_blending_z}
g_{\rm obs} &= \int dz_{\rm p}\, g(z_{\rm p}) \int d\bm{G}\, n_{\bm{G}}(z_{\rm p})\, R_{\rm self}(\bm{G}_{\rm p}) \notag \\
&\quad + \int_0^{\theta_{\rm max}} d\theta \int dz_{\rm s}\, g(z_{\rm s}) \int d\bm{G}_{\rm s}\, \tilde{N}_{\bm{G}}(z_{\rm s},\theta) \notag \\
&\qquad \times \int dz_{\rm p} \int d\bm{G}_{\rm p}\, n_{\bm{G}}(z_{\rm p})\, R_{\rm blend}(\bm{G}_{\rm p}, \bm{G}_{\rm s}, \theta). \; 
\end{align}

The first term is the integration of the true shear weighted by the `self-response'expressed by $R_{\rm self}$ over $z_{\rm p}$, which is the redshift of the primary galaxy. The second term represents the integration of the true shear over redshift, $z_{\rm s}$, weighted by both the blending response and the occurrence of blends, with the secondary galaxy in each pair located at that redshift. Given that the light of {undetectable}  objects can have an impact, $\tilde{N}_{\bm{G}}(z_{\rm s},\theta)$ is defined as the un-normalised multi-dimensional distribution per unit redshift and per unit sky area of the truly existing galaxies' properties at separation $\theta$ from the primary galaxy. The spatial density and radial distribution of neighbours are assumed to be purely geometric, as galaxies at different redshifts are positionally independent. Consequently, $\tilde{N}_{\bm{G}}(z_{\rm s}, \theta)d\theta \propto 2\pi \theta d\theta$. This assumption may break down under higher order effects such as magnification and foreground–background alignments. The blending response diminishes beyond a sufficiently large separation, $\theta_{\rm max}$, and strongly depends on the shape measurement procedure, including neighbour-removal algorithms. The blending response,
\begin{equation}
    R_{\rm blend}(\bm{G}_{\rm p},\bm{G}_{\rm s},\theta)=\frac{\delta e_{\rm p}}{g_{\rm s}(\theta)}
,\end{equation}
is defined to be the change in the measured ellipticity, $e_{\rm p}$, of the detection associated with the primary galaxy, $\bm{G}_{\rm p}$, at $z_{\rm p}$ per unit shear, $g_{\rm s}(\theta),$ applied to the secondary galaxy, $\bm{G}_{\rm s}$, at $z_{\rm s}$ that is separated from the primary galaxy by an angular distance, $\theta,$ on the sky. The self response, $R_{\rm self}$, is the change of the same quantity per unit shear $g_{\rm p}$ applied to the primary galaxy itself,
\begin{equation}
    R_{\rm self}(\bm{G}_{\rm p}) = \frac{\delta e_{\rm p}}{g_{\rm p}}.
\end{equation}
It accounts for the impact on the measurement of the sum of light of both the primary galaxy and the neighbouring galaxies, while the latter are not being sheared. We note that it is only in the case of isolated galaxies that we would expect $R_{\rm self}=R$ with the response $R$ that would be measured when the same shear is applied to the entire scene of galaxies. 

Both responses depend on the properties of the galaxies involved, their relative positions, the selection of detections, the shape measurement method, and observing conditions. While, in principle, the relative orientations of galaxies’ position angles and the shear direction influence the blending response, their statistical independence and isotropy render this dependence irrelevant over large samples, which we discuss in Sect. \ref{sec:measurements}.

The above formula sums up the responses of the detection's measured shape to individual secondary galaxies in cases where multiple neighbours are present. We treat the responses as if they were evaluated independently. This is strictly incorrect, as the presence of a third or more galaxies reduces the influence of any single secondary on the measured shape of the `blob'. However, the treatment of the effect greatly simplifies if one can work on galaxy pairs and take multiple-component blends as combinations of pairs. We find that such an approximation mostly holds and strongly deviates only when multiple bright neighbours are present, as discussed in Appendix \ref{sec:linearity}, which is rare.

In most current shear calibration approaches, the shear response of a blend is evaluated as a whole. Under the assumption of linearity described in the paragraph above, the total response can be expressed as 
\begin{equation}
R = R_{\rm self} + \sum_i R_{{\rm blend},i}, \;
\end{equation}
where $i$ is the index of the neighbouring galaxies. This $R$ is calibrated by simulations where a constant shear is applied to all galaxies. It is also the normalisation we assume that $n_{\gamma}$  takes (cf. Eq. ~\ref{eqn:ngamma}). Methods that do not rely on external simulations, such as \textsc{Metadetection}, also only estimate  $R$ by applying the same shear to the entire deconvolved image, rather than disentangling the contributions of individual components. Here,
\begin{equation}
    \delta e = R_{\rm self} \times g_p + \sum_i R_{{\rm blend}, i} \times g_{s,i} = R\times g_{\rm p} \; ,
\end{equation}
where the second equality holds only because in these simulations $g_{\rm p}=g_{\rm s}=g$ is the same for all galaxies.

These constant shear methods account for the impact of both the light of neighbouring galaxies and their associated shear. However, they neglect the variation in shear among the blending components arising from the redshift dependence of the shear field. Under this incorrect assumption, Eq. \ref{eq:g_blending_z} would be rewritten as

\begin{align}
\label{eq:g_blending}
g_{\rm obs}^{\rm const} &= \int dz_{\rm p}\, g(z_{\rm p}) \int d\bm{G}\, n_{\bm{G}}(z_{\rm p})\, R_{\rm self}(\bm{G}_{\rm p}) \notag \\
& + \int_0^{\theta_{\rm max}} d\theta \int dz_{\rm s}\, \int d\bm{G}_{\rm s}\, \tilde{N}_{\bm{G}}(z_{\rm s},\theta) \notag \\
&\quad \times \int dz_{\rm p}\, g(z_{\rm p})\int d\bm{G}_{\rm p}\, n_{\bm{G}}(z_{\rm p})\, R_{\rm blend}(\bm{G}_{\rm p}, \bm{G}_{\rm s}, \theta). \; 
\end{align}
We note that $g(z_{\rm p})$ in the blending term is now (erroneously) associated with the primary galaxy's redshift. 

Taking the difference between Eq. \ref{eq:g_blending_z} and Eq. \ref{eq:g_blending}, we find the correction on these traditional shear calibrations as
\begin{align}
g_{\rm obs} - g_{\rm obs}^{\rm const} 
&= \int dz\,dz'\,d\theta\,d\bm{G}_{\rm s}\,d\bm{G}_{\rm p}\, g(z) \notag \\
&\quad \times R_{\rm blend}(\bm{G}_{\rm p}, \bm{G}_{\rm s}, \theta) \notag\\
&\quad \times \left[ \tilde{N}_{\bm{G}}(z,\theta)\, n_{\bm{G}}(z') - \tilde{N}_{\bm{G}}(z',\theta)\, n_{\bm{G}}(z) \right] \notag\\
&\equiv \int dz\, g(z) \Delta n_{\gamma}(z) \; ,
\end{align}
where $z_{\rm p}$ and $z_{\rm s}$ are replaced with $z$ and $z'$, and are swapped when necessary to simplify the equation. The shift in effective redshift distribution is given as
\begin{align}
\label{eq:correction}
\Delta n_{\gamma}(z) &= \int dz'\,d\theta\,d\bm{G}_{\rm s}\,d\bm{G}_{\rm p}\,\notag \\
&\quad \times R_{\rm blend}(\bm{G}_{\rm p}, \bm{G}_{\rm s}, \theta) \notag \\
&\quad \times \left[ \tilde{N}_{\bm{G}}(z,\theta)\, n_{\bm{G}}(z') - \tilde{N}_{\bm{G}}(z',\theta)\, n_{\bm{G}}(z) \right] \; \\
& \equiv n^{\rm}_{\gamma,\rm blend}(z)-n^{\rm const}_{\gamma, \rm blend}(z).
\end{align}
The effect cancels in the case of constant shear, or where $z=z'$, as the blending of galaxies at the same redshift does not introduce mixing signals. Then, $n^{\rm}_{\gamma, \rm blend}(z)$ represents the effective redshift distribution accounting for blending in a redshift-varying shear field, whereas $n^{\rm const}_{\gamma,\rm blend}(z)$ corresponds to the `constant shear' case.

Equation~\ref{eq:correction} can be quantified by modelling each of the components. The true population $\tilde{N}_{\bm{G}}(z,\theta)$ needs to be modelled to account for the impact, especially of faint objects whose abundance is uncertain. We also need an emulator for $R_{\rm blend}(\bm{G}_{\rm p},\bm{G}_{\rm s},\theta)$. Because measured properties of the detection associated with the primary galaxy, such as flux and size, are often influenced by nearby objects and are conditioned on the separation between the primary and secondary galaxies, they become degenerate when quantifying the blending response. Besides, secondary galaxies often have no associated detection. This makes it difficult to operate with galaxy properties measured directly from data. Therefore, we establish the relationship between the blending response and the true properties, and estimate the shift in the effective redshift distribution using the input catalogue rather than the data itself. This necessitates a further step to translate the true population into observed samples $n_{\bm{G}}(z)$, including detection, selection, and the tomographic binning of samples. The emulation of these processes and the modelling of $R_{\rm blend}$ will depend on the specifications of the survey and the algorithms used on the actual data. These include the observing conditions, calibration and data reduction, detection and de-blending methods, and shape measurement, as well as redshift characterisation.

\begin{figure*}
    \centering
    \includegraphics[width=0.45\linewidth]{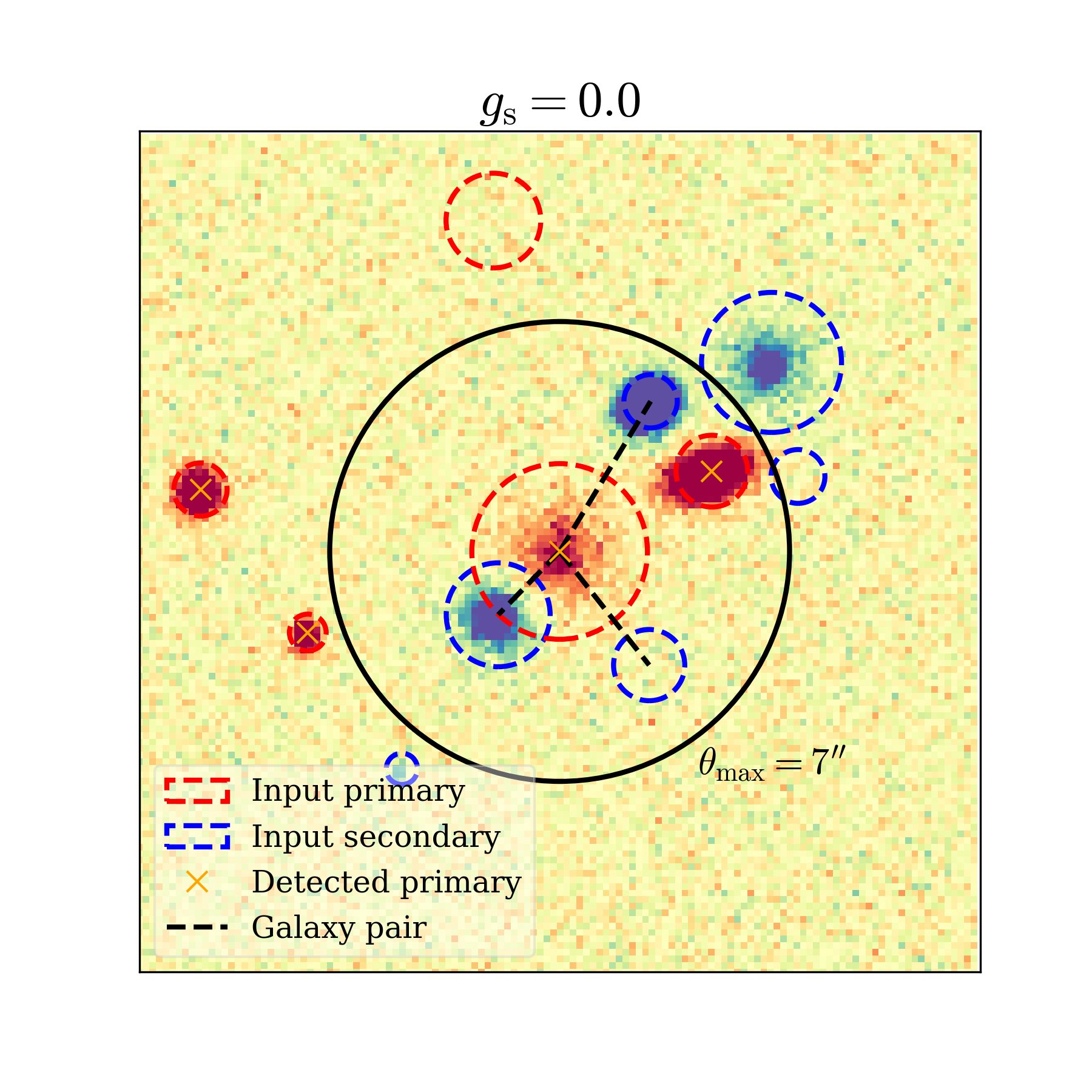}
    \includegraphics[width=0.45\linewidth]{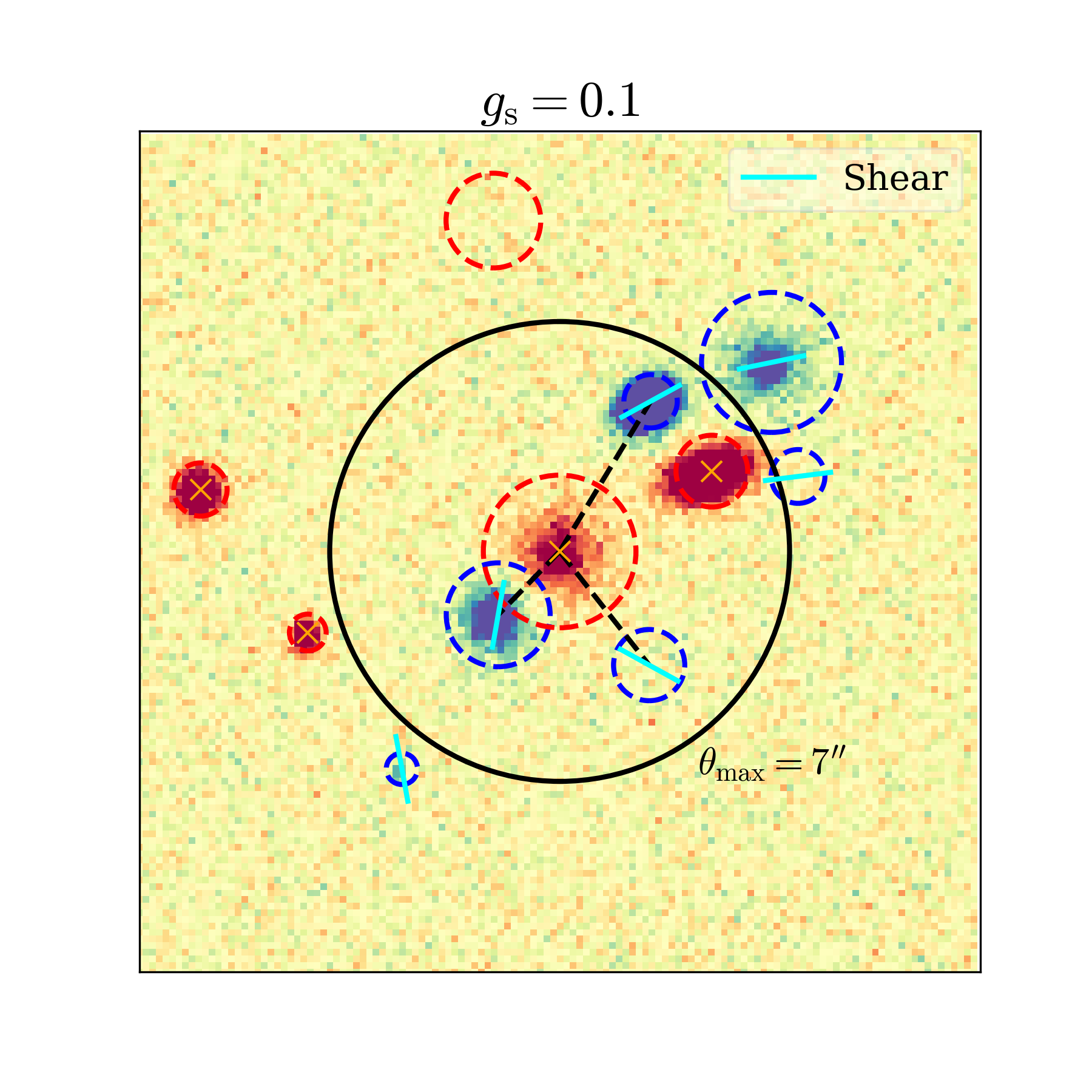}
    \caption{Illustration of the half-sheared simulation used to capture the blending shear response. Input secondary galaxies are marked with  dashed blue circles, and input primary galaxies with red ones. A subset of the primary galaxies is associated with detections as described in Sec.~\ref{sec:half_shear}, indicated by orange crosses. Two sets of simulations share the same galaxy catalogue and noise realisation, but differ in the shear applied to secondary galaxies: $g_{\rm s}=0.0$ (left panel) and $g_{\rm s}=0.1$ in uncorrelated directions (right panel). Galaxy pairs are identified by finding each secondary galaxy within $10$ arcsec of a primary detection found in both simulations. For each pair, the change in the measured shape of the primary galaxy's detection along the shear direction is calculated between the two simulations, which is used to estimate $R_{\rm blend}(\bm{G}_{\rm p}, \bm{G}_{\rm s}, \theta)$.}
    \label{fig:example}
\end{figure*}

\section{Simulating blending}

In this section, we first describe the process of generating the galaxy mock catalogue and how we use it to simulate blending-affected images. We then provide details on the procedure for retrieving $R_{\rm blend}$ from the simulations, as well as the shape measurement method we adopted. Additionally, we outline the organisation of the training datasets.

\subsection{Galaxy population and mock catalogue}
\label{population}

A realistic model describing the galaxy population is crucial to most shear calibration approaches due to the dependence of the shear bias on galaxy properties. We briefly describe the generation of our mock catalogues in four steps: 1) sampling galaxies' physical properties from a set of observationally informed priors provided in \textsc{Prospector}-{$\beta$} \citep{wangWangInferringMoreLess2023}; 2) generating spectral energy distributions (SEDs) and magnitudes based on these properties using stellar population synthesis code \citep{Luca2024Stellar,Conroy2009,Conroy2010}; 3) assigning morphological parameters to each sample utilising a learning model given its magnitude and redshift \citep{liKiDSLegacyCalibrationUnifying2023}; and 4) resampling the catalogue to match a observational calibrated stellar mass function (SMF; \citealt{lejaNewCensus022020}). We include more details in Appendix \ref{sec:catalogue_app}.

\subsection{Blending on a half-sheared sky}
\label{sec:half_shear}

We  reiterate that $R_{\rm blend}$ is fundamentally tied to the properties that could influence the observed ellipticity. In the presence of two nearby sources, a primary galaxy and a secondary galaxy, the relevant properties include parameters of both sources, such as their flux, flux concentration, size, axis ratio, and position angles relative to the direction of shear.

These same properties determine source detection. Typically, the information from the primary source alone suffices to determine the detection outcome. However, in certain cases, nearby objects also play a role. For example, a faint galaxy located near a bright counterpart may fail to be successfully de-blended, whereas it would normally be detected if isolated. As discussed in detail in Sect. \ref{p_model}, we construct our detection emulator using the complete blending pair information and compare its results to those obtained when only isolated galaxy properties are considered.

We designed a specialised simulation setup to extract $R_{\rm blend}$ individually, as is illustrated in Fig. \ref{fig:example}. Instead of manually placing galaxy pairs onto postage stamps, simulating galaxies within a realistic sky environment proved both more efficient and effective, as it produces a realistic ensemble of pairs and captures underlying effects inherent to an actual observation pipeline. To do this, we utilise the SKiLLS image simulation package \footnote{\url{https://github.com/KiDS-WL/MultiBand_ImSim}} developed for the KiDS-Legacy analysis \citep{liKiDSLegacyCalibrationUnifying2023}. SKiLLS integrates the modules necessary for simulating weak lensing, including multi-band image generation, exposure stacking, object identification based on Source Extractor \citep{Bertin1996}, and optional internal measurement of galaxy shape and redshift. 

With SKiLLS, we can generate images with a customised input catalogue and survey specifications. To isolate $R_{\rm blend}$ from $R_{\rm self}$, the primary galaxy must remain unsheared. Therefore, we simulated a sky in which a random half of the galaxies were sheared by $10\%$ in randomised directions, while the other half remained unsheared. The unsheared galaxies were designated as primaries, and the sheared ones as secondaries, with their indices recorded in the input catalogue. While 10\% shear is typically a large value in terms of cosmic shear, it is a good approximation considering the response of a detection's shape to a secondary galaxy's shear is typically small, $R_{\rm blend} \ll 1$, as we show in Appendix~\ref{sec:linearity}. 

To create a baseline for comparison, an identical sky was also simulated, preserving the same noise realisation and input galaxy catalogue, but with no applied shear. Both suites underwent the same detection and shape measurement pipeline, producing a `positive' catalogue (with $g_{\rm s}=0.1$) and a `negative' catalogue (with $g_{\rm s}=0.0$) containing both detections and primary shape measurements. Then, $R_{\rm blend}$ for each detected primary galaxy and one of its companions was estimated as the difference in the measured ellipticities between the two catalogues, expressed as $\delta \textbf{e}$, divided by $g_{\rm s} = 0.1$.

It is important to define how a detection is associated with an input primary galaxy. When blending occurs, this association is ambiguous. However, an operation can be specified that uniquely determines the primary galaxy for each detection and this operation needs to be used consistently both in the redshift calibration step that yields the initial $n(z)$ and in determining the blending correction \citep[e.g.][]{Everett2022,MylesDESredshift2021,maccrannY3ResultsBlending2022}. In this work, an input galaxy is assigned to be the true source associated with a detection if its centroid lies within 2.5 pixels of that of the detection. If multiple input galaxies fall within this radius, the one whose magnitude is closest to the measured magnitude of the detection is assigned as the source. Since the measured flux typically reflects the combined light of blended sources, this criterion often results in the brighter galaxy being identified as the source. Consequently, fewer pairs are found where a faint primary is blended with a bright secondary, as further discussed in Sect.~\ref{p_model}. This introduces a selection effect in the mean measured $R_{\rm blend}$ across pairs, though it does not mean the setup of our simulation introduces extra bias. Since ambiguous blends are typically interpreted as the most impactful galaxy whose measurements are affected by other galaxies in the vicinity, this tendency in our setup to prefer pairs with bright primary and faint secondary of ambiguous blends reflects observational reality.

The positive and negative detection catalogues are not necessarily identical, primarily due to stochastic effects and shear-dependent detections. We take the intersections of the two catalogues and search for the secondary galaxies for each detection. This is achieved by employing a $k$-dimensional tree to identify input galaxies within the secondary division based on sky position and within a fixed separation. Since only detections present in both catalogues are used to identify pairs, the sample excludes blends that may be subject to detection biases induced by the shear of secondary galaxies. Shear-related detection bias is primarily sensitive to the relative orientations between the shapes of the primary and secondary galaxies and the direction of the applied shear. As a result, the intersection sample may omit certain combinations of these relative angles, potentially introducing a selection bias. However, we expect this effect to be minor, as we do not observe a significant dependence of the mean blending shear response on these angles.

For each detection associated with a primary, we look for its neighbouring secondary galaxies within an angular separation $\theta_{\rm max}$. We adopt $\theta_{\rm max}=7~\rm arcsec$ to ensure it includes all impactful neighbours. As shown in the last panel of Fig. \ref{fig:delta_e_1}, the blending shear response of galaxy pairs at $\theta_{\rm max}=7~\rm arcsec$ is observed to be negligible.

We randomise the shear directions of each secondary galaxy as a deliberate design to isolate their individual contributions when multiple secondary galaxies are nearby a primary. For each identified pair, we compute two components of $\delta \textbf{e}$: the component along/perpendicular to the shear direction, $\delta e_{\rm +}$, and the component at $\pm 45^{\circ}$, $\delta e_{\times}$. For a single sample, $\delta e_{\times}$ is not necessarily zero when the secondary galaxy is positioned off the shear direction relative to the primary galaxy. We can assume the secondary galaxies are isotropically distributed around the primary. The same applies to their orientation angles. Consequently, when averaging over many samples, we obtain $\langle \delta e_+ \rangle = R_{\rm blend} \times g_{\rm s}$ and $\langle \delta e_{\times} \rangle = 0$. By taking the $\delta e_{+}$ for each galaxy pair, we isolate the contribution from the specific secondary galaxy in that pair, while ensuring that the expected contribution from all other secondary galaxies within the aperture averages to zero in the direction of interest. Nevertheless, the presence of these additional secondary galaxies introduces extra noise in the measurements. Ultimately, these steps result in the creation of a final measurement catalogue. This catalogue contains the input properties of each galaxy pair, as well as the differences in measured shapes, $\delta e_{\rm +}$ and $\delta e_{\times}$.

Using a similar procedure for investigating and predicting the detection outcome, we generated a catalogue that includes the input properties of each galaxy pair along with a boolean value indicating whether the primary is associated with a detection. Here, we used the catalogue based on the unsheared realisation of the image to avoid shear-dependent detections. This time, instead of
considering all companions within a $3~\rm arcsec$ aperture of each source, we limited the search to the closest neighbour within a maximum separation of $3~\rm arcsec$. While we acknowledge that the most significant impact on a source's detection does not always originate from the nearest neighbour but rather from a combination of distance and the neighbour's appearance, this simplification is sufficient for a first-order analysis. As noted earlier, the detection catalogue was constructed solely from the null suite, disregarding the effects of shear. For sources that lack a companion within the specified separation, a Null neighbour was assigned in the catalogue.

We adopted a constant, round Moffat profile as the PSF model, characterised by $\beta = 2.4$ and a ${\rm FWHM} = 0.6~\rm arcsec$. The pixel scale is $r_{\rm pixel} = 0.2~\rm arcsec$ on a side. In principle, variations in the PSF across the focal plane and its anisotropies are critical factors in shape measurement. The impact of PSF ellipticity on blending response determination is discussed in detail in Appendix \ref{sec:PSF}. Additionally, we included constant Gaussian noise with $RMS=6$ $\mathrm{counts}$ per pixel and a zero background level, with a zero-point magnitude of $30$. We simulated $r$ band images that include all galaxies with magnitudes brighter than 27. The detection limit of this setup is $\sim 26$th magnitude at $5 \sigma$, comparable to the observing conditions of HSC \citep{Aihara2022}. The \textsc{SExtractor} configuration used in source detection is shown in Table \ref{tab:sextractor_config}.

\begin{table}
\centering
\caption{\textsc{SExtractor} configuration parameters used for source detections.}
\begin{tabular}{ll}
\hline
\addlinespace[0.5ex]
Parameter & Value \\
\hline 
\addlinespace[0.5ex]
\texttt{DETECT\_MINAREA}  & \texttt{4} \\
\texttt{DETECT\_THRESH}   & \texttt{1.5} \\
\texttt{ANALYSIS\_THRESH} & \texttt{1.5} \\
\texttt{DEBLEND\_NTHRESH} & \texttt{32} \\
\texttt{DEBLEND\_MINCONT} & \texttt{0.001} \\
\texttt{CLEAN}            & \texttt{Y} \\
\texttt{CLEAN\_PARAM}     & \texttt{3.0} \\
\texttt{MASK\_TYPE}       & \texttt{CORRECT} \\
\texttt{BACK\_TYPE}       & \texttt{MANUAL} \\
\texttt{BACK\_VALUE}      & \texttt{0.0,0.0} \\
\hline
\end{tabular}
\label{tab:sextractor_config}
\end{table}

\subsection{Training data and validation sets}

With one square degree as a tile, each set is defined as two tiles, one with half of the input sources sheared and a null suite. Each set draws galaxies from a four-square-degree input catalogue based on our fiducial population model with different random states and shuffled positions to avoid duplicated pairs. We simulated 200 sets as our training data and 200 sets as the test data, each generating a response and a detection catalogue. A selection on the catalogues is then applied, retaining galaxies with input magnitudes between 18 and 26, and input effective radius $r_e$ between $0.1~\rm arcsec$ and $1.5~\rm arcsec$. The galaxies outside this range are either too rare to have an impact or too faint to have a relevant blending response.

\subsection{Shape measurement}
\label{shape}

The impact of blending depends on the specific methods used in shape measurement, which vary among existing surveys. Masks are frequently employed to be insensitive to neighbouring galaxies \citep{jarvisScienceVerificationWeak2016,zuntzDarkEnergySurvey2018}. Additionally, galaxies with nearby companions within a certain contamination separation are often rejected, resulting in a loss of samples. To address this, more advanced approaches have been developed to avoid the significant reduction in galaxy density caused by these rejections. For example, de-blending algorithms can be applied to assign flux, separating the contribution of blended detections at the pixel level (i.e. \citealt{drlica-wagnerDarkEnergySurvey2018}). Masking and flux assignment, when used together, can potentially mitigate the effects of blending. but only if the blends are correctly identified. We do not apply any pixel-level masking or flux assignment in our shape measurement process. Additionally, background misestimation, particularly due to local light from very faint objects, has been shown to affect measured shapes after subtraction \citep{hoekstraStudySensitivityShape2017}. While our fiducial setup does not include background subtraction, we acknowledge that it plays an important role in quantifying blending. A more uniform background subtraction is expected to have a smaller impact on the determination of $R_{\rm blend}$, although it neglects local sky variations. Conversely, localised background subtraction could affect $R_{\rm blend}$, as blended light may be incorrectly treated as background. For emulation purposes, it is essential to replicate these treatments used in the targeted data as closely as possible.

We utilised the model-fitting procedure provided by the NGMIX package\footnote{\url{https://github.com/esheldon/ngmix}} \citep{sheldonPracticalWeaklensingShear2017} and the Galsim.HSM package\footnote{\url{https://github.com/GalSim-developers/GalSim}} \citep{Rowe2015Galsim} for initial galaxy size estimation. The observed galaxy is first fitted using the \texttt{Galsim.HSM.gauss\_mom} function to obtain an initial size estimate \citep{guinotShapePipeNewShape2022}. The pre-PSF ellipticity was then determined by fitting the object with a single Gaussian model, convolved with the PSF Gaussian, based on the Galsim-derived size guess. Although more sophisticated parametric profiles are available in NGMIX, the Gaussian profile is generally sufficient and computationally more efficient. We used the true PSF input from the simulations rather than a reconstructed PSF. While PSF modelling errors are typically a critical factor in shear estimation (e.g. Sect. 5.3 in \citealt{liKiDSLegacyCalibrationUnifying2023}) and depend on both the survey and the measurement method, we treat them as a secondary effect in the context of quantifying the blending response.

\section{Emulating secondary shear response}

\begin{figure*}
    \centering
    \includegraphics[width=0.9\linewidth]{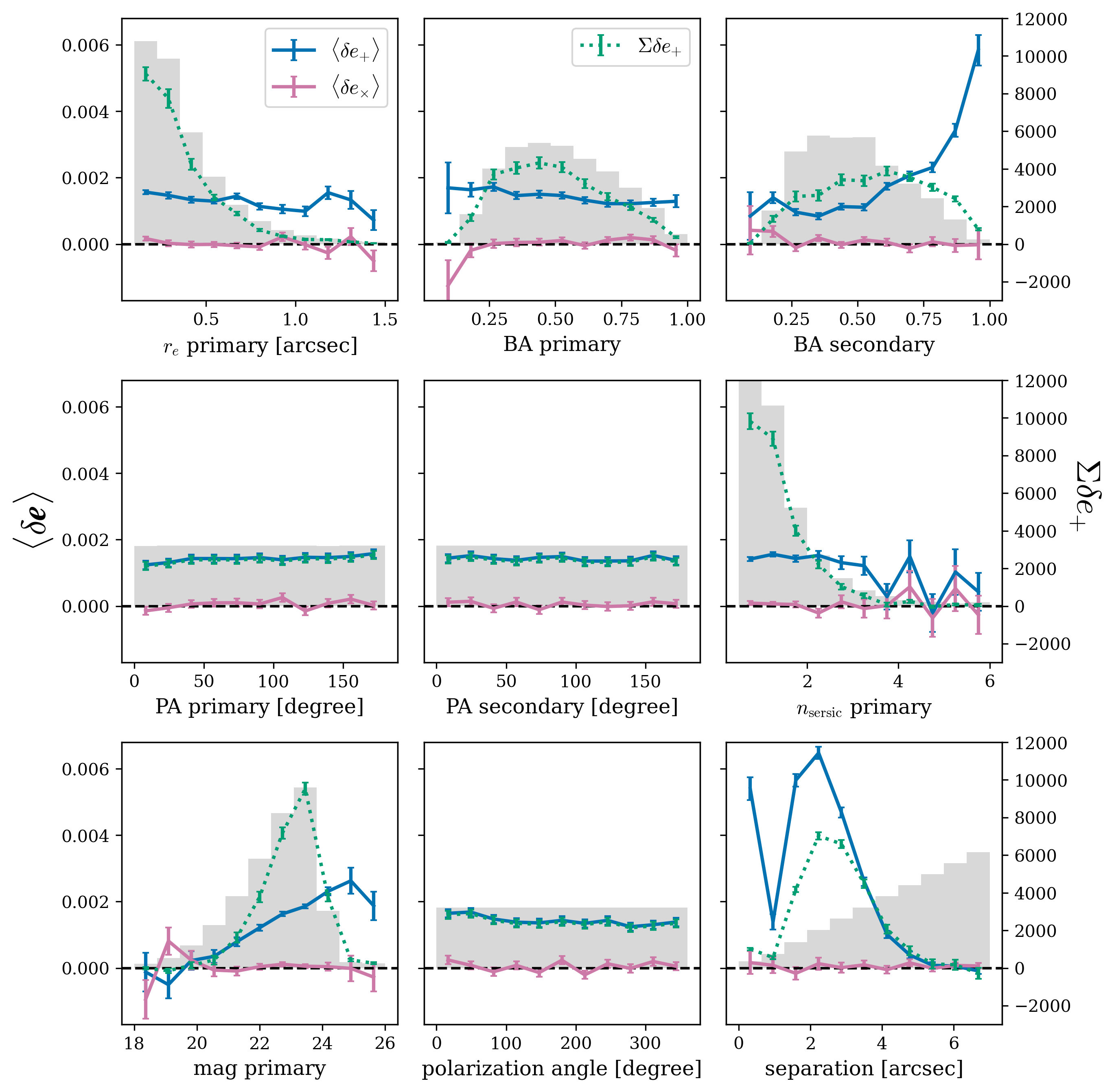}
    \caption{Binned average of $\langle \delta \bm{e} \rangle$ versus some of the properties of each blend, including effective radius ($r_e$), axis ratio (BA), position angle (PA), Sersic index ($n_{\rm sersic}$), magnitude (mag), the polarisation angle between the pair, and their separation on the sky. Dependence on properties that closely correlate with the secondary galaxy's brightness is shown in Fig. \ref{fig:delta_e_2}. The blue curves show the component, $\langle \delta e_{\rm +} \rangle$, projected in the direction of the shear applied to the secondary, while the pink ones are in the direction 45 degrees to the shear, $\langle \delta e_{\times} \rangle$. Normalised histograms of each property are plotted as filled histograms. The {population-weighted} $\Sigma \delta e_{\rm +}$ are shown as green curves.}
    \label{fig:delta_e_1}
\end{figure*}

In this section, we describe the construction of blending response and detection probability emulators using machine learning algorithms. We begin in Sect. \ref{sec:measurements} by presenting the $R_{\rm blend}$ measurements obtained from simulations and examining their dependence on the properties of blends. Next, we build an emulator based on these measurements using a boosted tree model in Sect. \ref{xgboost}, where we briefly outline the training and tuning procedures of the models and compare their predictions with measurements from simulations.  We also evaluate the robustness of the response model to the uncertainties in galaxy populations in Sect. \ref{population}.

\subsection{$R_{\rm blend}$ versus galaxy properties}
\label{sec:measurements}

We first analysed the measured shifts in the ellipticities of detections when shearing their secondary galaxies. Fig. \ref{fig:delta_e_1} presents the binned average $\delta \bm{e}$ of detection relative to the relevant parameters of the associated galaxy pair. The component $\delta e_{\times}$, at 45 degrees relative to the shear direction, is consistently observed to be zero on average over many pairs. Several forms of isotropy cancel the cross component:

1. Isotropy of secondary galaxies:\ A secondary galaxy that is offset oriented with respect to the shear direction can introduce a cross component in the measured shape distortion, as the blob is not spin-2 symmetric. However, if the orientations of secondary galaxies are uniformly distributed, the average $\delta \bm{e}$ over many realisations reflects the response to effectively round secondary galaxies due to this random spinning, and the cross component vanishes. Since we always project shapes with respect to the shear direction, the relative polarisation angle between the primary and secondary galaxy can be assumed to be uniformly distributed and, hence, it should not affect the outcome on average.

2. Isotropy of the shear field:
Multiple secondary galaxies may exist around a single primary galaxy. The measured shape responds to the combined shear from all surrounding secondaries. However, since these shears in our simulation are distributed randomly in direction, for each pair of interest, the shears of extra secondary galaxies contribute as noise over large samples. This applies to both components of $\delta \bm{e}$.

3. Isotropy of primary galaxy orientations:\ Under the previous assumptions, the $\delta \bm{e}$ value with respect to the secondary galaxy's shear signal is independent of the primary galaxy's intrinsic orientation, as the summed light of the two galaxies follows spin-2 symmetry. However, higher order effects during the fitting process could break such symmetry, for example, when the centroid of the fitted profile is a free parameter. These effects would still cancel and lead to $\langle \delta e_{\times} \rangle =0$ under the assumption that primary galaxies are randomly oriented.

Most of the parameters shown in Fig.~\ref{fig:delta_e_1} exhibit only a weak influence on the average tangential component, including the position angles (PA) and polarisation angles of the member galaxies. Primary galaxy properties such as effective radius ($r_e$), axis ratio (BA), and Sérsic index ($n_{\rm sersic}$) appear to have little relevance. In contrast, the primary galaxy's magnitude has a clear impact: very bright primaries (with magnitudes below 20) are barely affected by blending, while the mean effect becomes stronger as the primaries get fainter. The angular separation shows an intuitive trend: the closer the two galaxies, the stronger the blending effect. A {notable}  dip in $\delta e_{\rm +}$ occurs around 1 arcsec. This can be explained as follows: when galaxies are very close, they often become unresolved, and the detection is assigned to the brighter component. As a result, fewer faint-primary/bright-secondary pairs are included, which reduces the average response. This highlights the influence of neighbouring galaxies on the detection process and provides motivation for incorporating blending information in our detection emulation, as discussed in Sect.~\ref{p_model}. At the same time, the non-linear increase in response with decreasing separation counteracts this effect, leading to a slight rise in $\delta e_{\rm +}$ at even smaller separations.

The {population-weighted} results, $\sum \delta e_{\rm +}$, are the summation of each found pair's $\delta e_{\rm +}$ and are plotted to illustrate the overall impact of different galaxy configurations. Since most parameters have a relatively minor influence individually, their {weighted} effect largely reflects their distribution, underscoring the importance of accurately constraining the galaxy population. Interestingly, $\sum \delta e_{\rm +}$ peaks at a separation of approximately $2~\rm arcsec$, while it remains relatively small at separations below $1~\rm arcsec$. This suggests that simply masking distant neighbours without much overlapping flux may be insufficient, and emphasises the need to properly account for nearby sources with overlapping pixels.

\begin{figure*}
    \centering
    \includegraphics[width=0.9\linewidth]{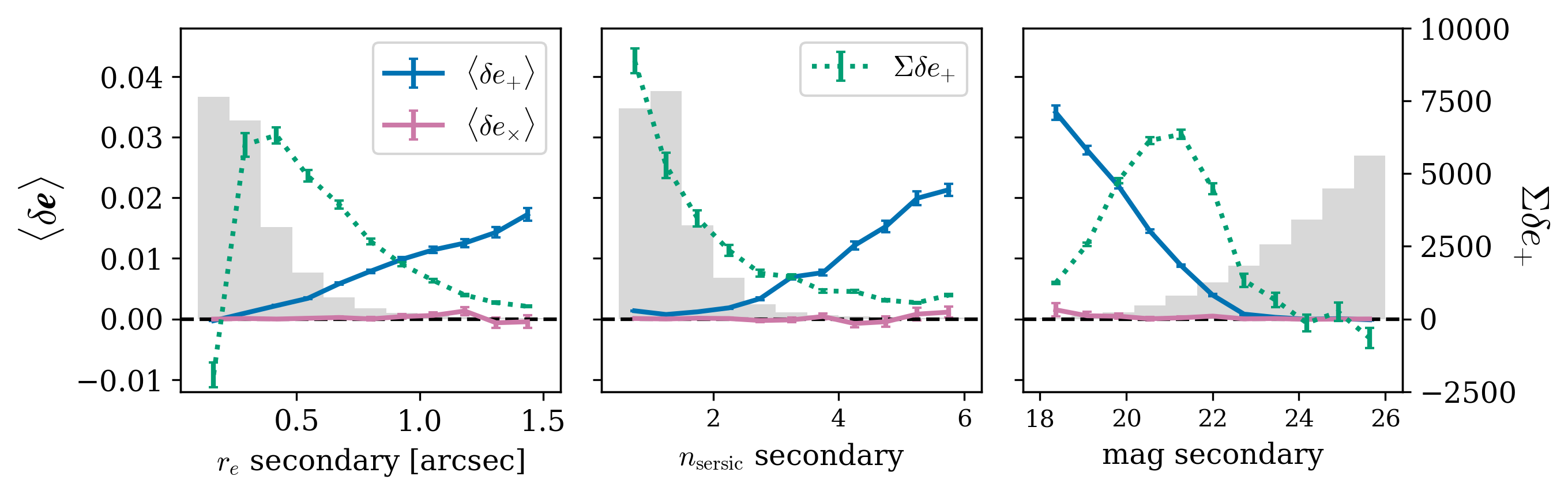}
    \caption{Same as Fig. \ref{fig:delta_e_1}, but showing the dependence of $\delta \bm{e}$ on properties of the secondary galaxy that closely correlate with its apparent brightness, with expanded vertical axis.}
    \label{fig:delta_e_2}
\end{figure*}

The impact of secondary size, $n_{\rm sersic}$, and magnitude is shown separately in Fig.~\ref{fig:delta_e_2}, where they exhibit a dominant influence on $\delta \bar{e}_{\rm t}$ at the per cent level. In general, larger, brighter, and more diffuse secondary galaxies have a stronger effect. The weighted results are determined jointly by the average influence and the population distribution. Overall, the strongest {population-weighted} impact is seen from small, bright neighbours.

It is worth noting that the trends shown above marginalise the remaining properties. As a result, even $\delta \bar{e}_{\rm t}$ implicitly encodes their underlying distributions. In reality, the joint influence of multiple properties reveals more complex relationships. While the marginalised results are useful for qualitative discussions, this complexity motivates the use of ML to capture the high-dimensional dependencies.

\subsection{Training boosted trees}
\label{xgboost}

The emulation task is based on well-structured datasets, where tree-based models are typically more suitable and cost-effective than deep learning methods, requiring minimal preprocessing. Accordingly, we build our $R$ and detection emulators with XGBoost, a scalable open-source library of a regularising gradient boosting framework for decision trees\footnote{\url{https://github.com/dmlc/xgboost}} \citep{chenXGBoostScalableTree2016}. XGBoost has a set of hyperparameters that must be determined prior to optimisation. These include the regularisation parameters $\gamma$ and $\lambda$, parameters describing the trees like the maximum depth and the minimum number of samples in a leaf, the learning rate, and the column subsample rate. The `learning rate' in the context of XGBoost is an analogy to the concept widely used in gradient descent, also known as shrinkage, which scales the tree value by a fraction to slow the convergence and avoid over-fitting \citep{chenXGBoostScalableTree2016}. We briefly introduce these parameters and review the main algorithm for training a boosted tree model in Appendix \ref{sec:trees_math}.

Finding the optimal hyperparameters is important in terms of model performance. We use Optuna for the tuning, an automatic hyperparameter optimisation software \footnote{\url{https://github.com/optuna/optuna}} \citep{akibaOptunaNextgenerationHyperparameter2019}. To efficiently search the parameter space, we adopt the Tree-structured Parzen Estimator (TPE) algorithm \citep{bergstraAlgorithmsHyperParameterOptimization2011, watanabeTreeStructuredParzenEstimator2023} built within Optuna. The search starts with 20 initial random samples of hyperparameters, and then TPE recursively samples new parameters from the learned posterior density that maximises the expected improvement over the current best trial. multi-variate TPE is also used to capture the interdependencies among parameters. 

The training data were randomly divided at a ratio of 1:4 into a validation set and a training set. The optimal hyperparameter set is determined by finding the best metric on the validation set, while penalising an overfitting, expressed as
\begin{equation}
    L = L_{\rm valid} - \alpha |L_{\rm train}-L_{\rm valid}|, \; 
\end{equation}
where $L$ is the objective function and $\alpha$ is set to 0.6. We built two separate models for shear response, $R$, and source detection probability, $p$. For the former, we chose the coefficient of determination for $L$,
\begin{equation}
    L = 1 - \frac{\sum_{i=1}^{n} (y_i - \hat{y}_i)^2}{\sum_{i=1}^{n} (y_i - \bar{y})^2},
\end{equation}
where $y_i$ is the true label from measurements and $\hat{y}_i$ is the predicted value. The latter is optimised using LogLoss,
\begin{equation}
    L = -\frac{1}{n} \sum_{i=1}^{n} \left[ y_i \log(\hat{y}_i) + (1 - y_i) \log(1 - \hat{y}_i) \right],
\end{equation}
where $y_i \in \{0, 1\}$ and $y_i \in [0, 1]$.

\subsection{$R_{\rm blend}$ regression}
\label{r_model}

\begin{figure*}
    \centering
    \includegraphics[width=0.9\linewidth]{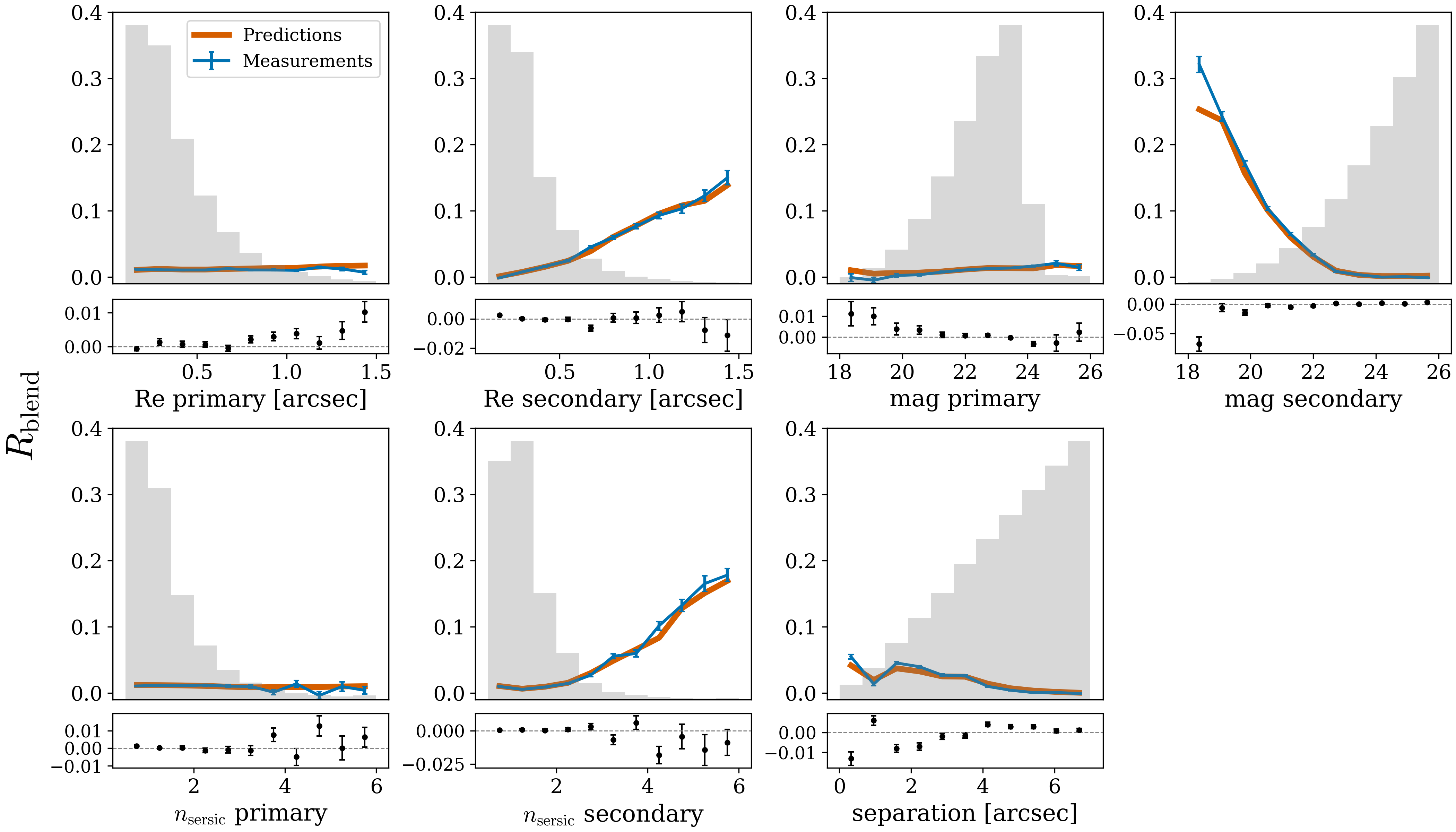}
    \caption{Binned average of blending response versus galaxy properties used in training the emulator. Blue curves with errors are measurements, while red curves are the predictions. Mini panels present the residuals between predictions and measurements with a varying y-axis scale. Normalised histograms of each property are plotted in grey histograms.}
    \label{fig:R_pred}
\end{figure*}

We began by training a regression model using the response catalogue of galaxy pairs generated from our simulations. We selected the most relevant ones based on one-dimensional (1D) averaged $\delta e_{\rm +}$ measurements under the aforementioned isotropy of blending configurations. These include the effective radii, magnitudes, and Sérsic indices of both the primary and secondary galaxies, as well as their angular separation. We neglected the dependence on galaxies' axis ratio as it has a relatively minor impact on the {population-weighted} results. The dependence on galaxy orientations and polarisation angle was not modelled, since it ought to be marginalised under the aforementioned isotropy assumptions. We used the true input values instead of measurements of these quantities for the considerations discussed in Sect. \ref{formula}.

In principle, $R_{\rm blend}$ depends on the observing conditions, with key factors including PSF, noise level, and pixel resolution. These elements modify the observed image on which shapes are measured and, consequently, influence the impact of the secondary galaxy's shear. However, since these observational effects primarily affect the image in a linear manner, we can define certain effective image properties (incorporating observing conditions) that render $R_{\rm blend}$ approximately invariant.

For example, a larger PSF increases the post-seeing size of galaxies, thereby enhancing blending. At the same time, it also tends to reduce the influence of shear, which acts at the pre-seeing stage. However, when rescaling the sizes and distances of the primary and secondary galaxies proportionately to the PSF,  $R_{\rm blend}$ should remain approximately unchanged. Likewise, background noise can influence $R_{\rm blend}$ by affecting the signal-to-noise ratio (S/N) of the primary and secondary galaxies. We expect the shear response to remain invariant when both the background noise level and the observed flux associated with a galaxy are scaled by the same factor.

In order to improve the model’s generalizability across spatial variations within a survey, or even across different surveys, and across different proposed galaxy populations, it is useful to construct the model using such rescaled properties. The parameters we use as inputs to the model are thus defined in terms of these normalised quantities, expressed as

\begin{equation}
\label{eq:scale1}
    \tilde{r}_e = \frac{r_e}{\sqrt{r_{\rm PSF}^2+r_e^2}},
\end{equation}

\begin{equation}
\label{eq:scale2}
    \tilde{\theta} = \frac{\theta}{\tilde{r}_{e,\rm primary}},
\end{equation}

\begin{equation}
\label{eq:scale3}
    \tilde{f} = \frac{f}{RMS*2\pi(r_{\rm PSF}/r_{\rm pixel})^2},
\end{equation}

where $r_{\rm PSF}$ is the effective radius of the galaxy in unit of $\mathrm{arcsec}$ and $f$ is the flux, which is converted into a magnitude after this scaling. Scaled values are denoted with a tilde. In Appendix \ref{sec:invariance}, we show how we verified that both $R_{\rm blend}$ and galaxy detection can be aptly generalised when using these definitions by comparing model predictions with simulations under varying observing conditions. However, we find that detections with very bright nearby neighbours introduce bias and, therefore, we excluded them from the remaining analysis, as detailed in Appendix \ref{sec:invariance}. The results in the paper present the pre-scaled values unless otherwise specified.

We applied the best-performing model to the test data and compared its predictions with the actual measurements, as illustrated in Fig. \ref{fig:R_pred}. The emulator's results generally align well with the measurements, particularly in regions with a higher density of data points. Conversely, the predictions are less accurate in areas where the training data are sparse. For example, the predictions for extremely bright secondary galaxies (with magnitudes $\lesssim 19$) deviate from the measured values by approximately $\sim 2\sigma$. That said, this discrepancy is likely to have a minimal impact due to the rarity of such sources. The overall average response $\bar{R}_{\rm blend}$ predicted by the emulator is $(1.114 \pm 0.0014)\times10^{-2}$ compared to the measured $(1.0790 \pm 0.0375)\times10^{-2})$. They are largely consistent within about $1\sigma$. The much smaller emulator error compared to the data error arises because each prediction corresponds to the averaged output of training data points that follow the same trajectory through the decision trees. As a result, the prediction catalogue often contains duplicate values, reflecting the fact that the emulator smooths over the intrinsic stochasticity present in the measured data, such as image noise and variations in nuisance parameters like the angles within galaxy pairs.

\begin{figure*}
    \centering
    \includegraphics[width=0.33\linewidth]{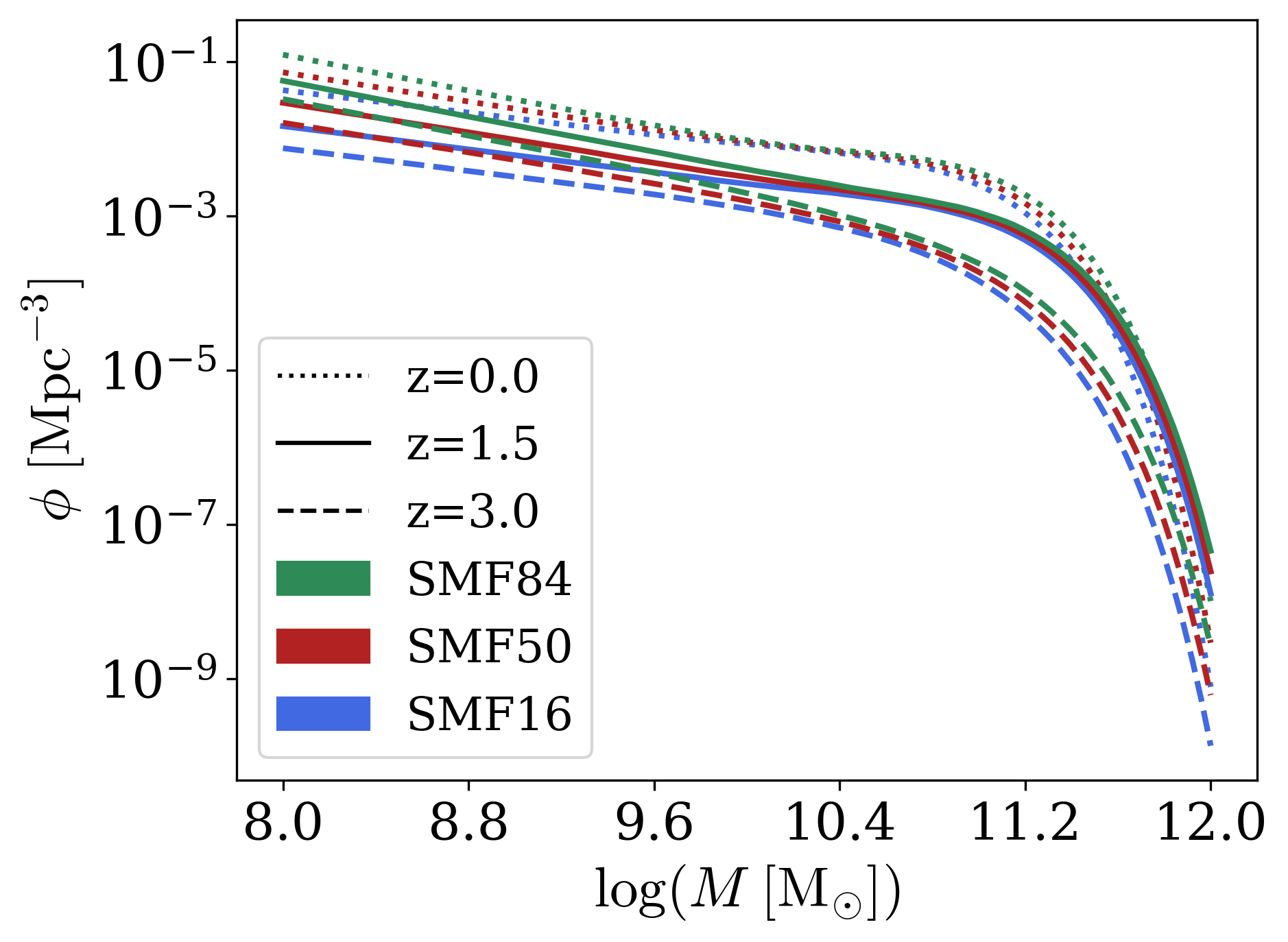}
    \includegraphics[width=0.33\linewidth]{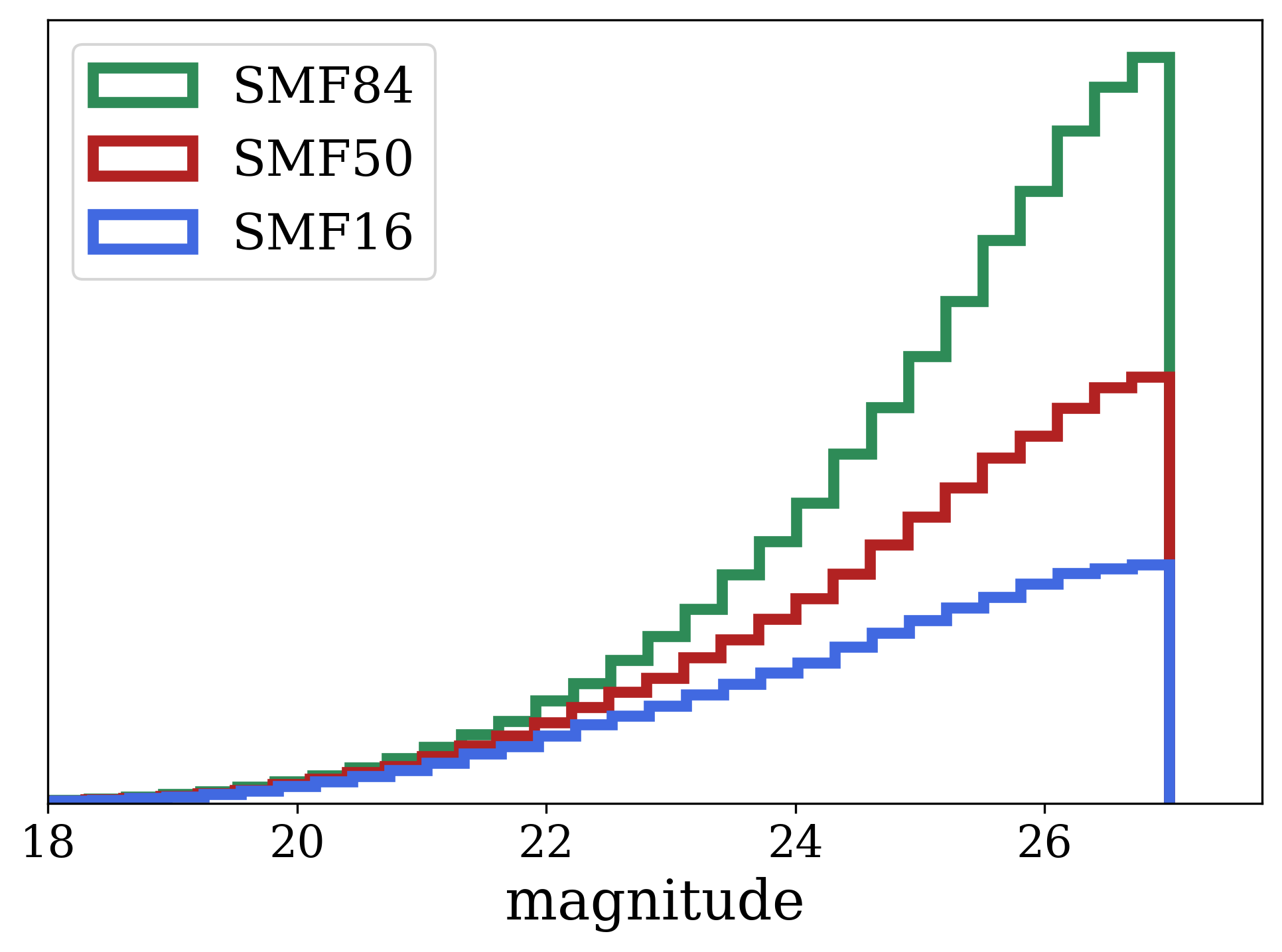}
    \includegraphics[width=0.33\linewidth]{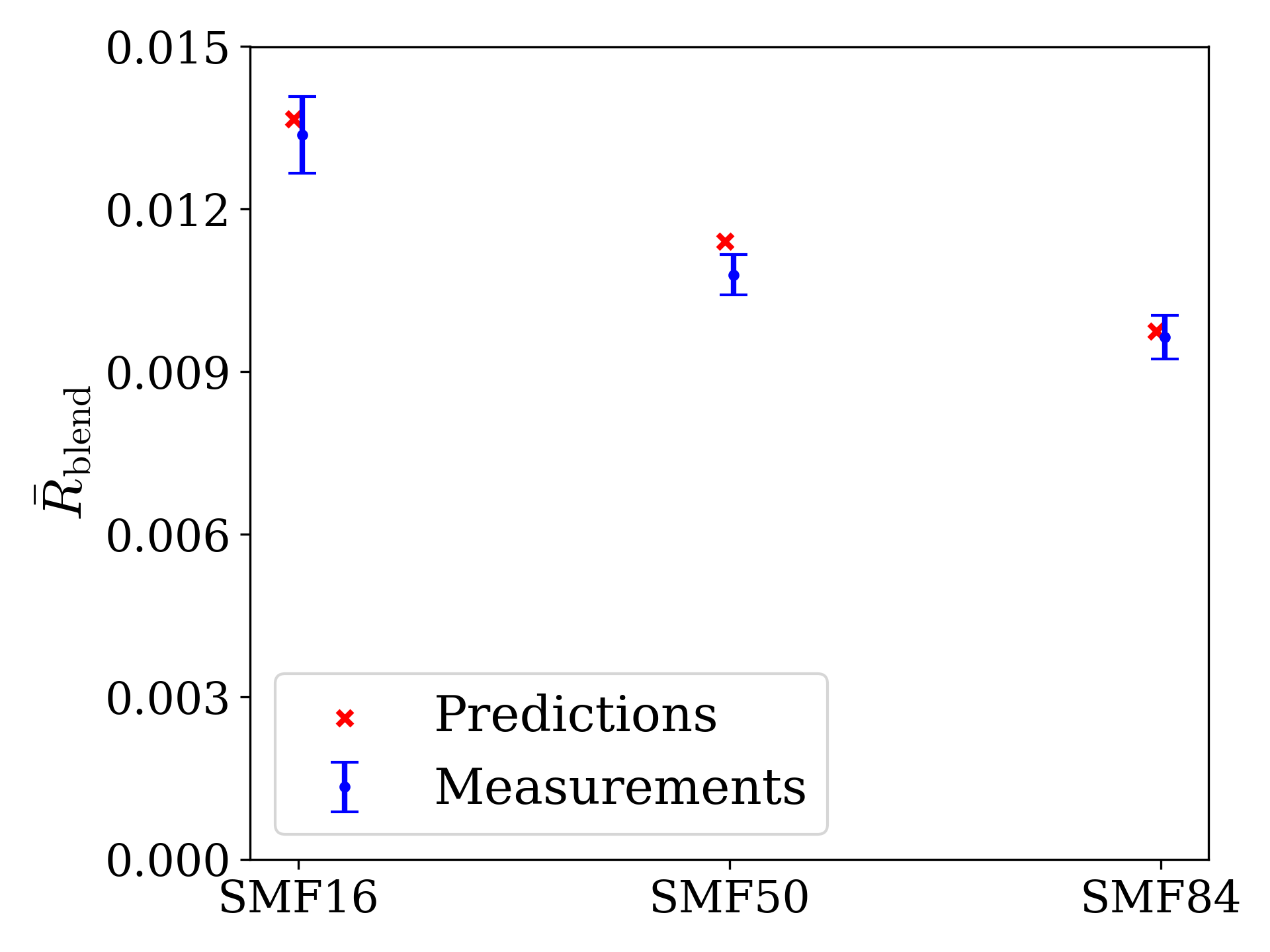}
    \caption{\textit{Left panel}: Stellar mass functions at $z=0.0,1.5,3.0$ using the \citet{lejaNewCensus022020} continuity model in dotted, solid, and dashed lines respectively. The model parameters and their uncertainties are taken from the same literature. We take the median model (SMF50, red lines) and models at $68\%$ confidence level, where SMF84 (green lines) denotes the upper limit and SMF16 (blue lines) denotes the lower limit. \textit{Middle panel}: Corresponding histograms of galaxy magnitude for different population models. \textit{Right panel}: Three test sets of simulations are generated using the models. The blue line with error bars presents the ensemble-averaged $R$ measured in the three sets, while the orange line shows the emulator predictions. The emulator is trained on a training set of SMF50 only.}
    \label{fig:pop_pred}
\end{figure*}

In practice, an emulator's performance can be influenced by the distribution of its training data and may degrade when applied to a test set with a different distribution. This issue is commonly known as the domain shift problem. Our response emulator was trained on simulations where galaxies were sampled based on SMF50, the median value derived from an observationally fitted SMF model, which might be biased from the actual distribution. Similar issues arise with varying observing conditions, which alter the data domain under the scaling as in Eqs. \ref{eq:scale1}-\ref{eq:scale3}, which we discuss in Appendix \ref{sec:invariance}. To assess whether the emulator performs adequately across diverse galaxy populations, we evaluated its predictions on the 1-$\sigma$ catalogues, SMF16 and SMF84, and compared them with simulation measurements, as shown in Fig. \ref{fig:pop_pred}. These uncertainties reveal a significant impact on the averaged $R$, with a $28\%$ decrease from SMF16 to SMF84. The response emulator successfully captures this variation, demonstrating its robustness in handling varying populations.

\subsection{Detection emulator conditioned on neighbouring galaxy}
\label{p_model}

The fact that we built our model of response upon galaxies' true properties has made it necessary to emulate the process of how observed galaxy samples are selected. In this work, we built a classification model that identifies input objects as detections, assigned to bins based on their redshifts. We assume we know the true redshifts for the tomographic tests below, while  we also need to ultimately emulate the bin assignment. 

Detecting and separating objects is a fundamental step in photometric analysis. Typically, an object is identified when a minimum number of adjacent pixels exhibit flux exceeding a predefined threshold. In scenarios involving multiple overlapping sources, algorithms like \textsc{SExtractor} de-blend them by locating branches where their number of levels surpasses a specified fraction of the pre-defined total number of levels. This often results in failure to identify objects given a very bright and close companion in the area. To capture this, it is essential to incorporate information about each source's neighbouring galaxies in addition to the source's own properties. A good approximation can be made that a single neighbour dominates the impact, mostly depending on some joint value of its distance and magnitude. We test two simplest cases: one assuming the most impactful neighbour to be the closest and the other to be the brightest one within a distance of 3 arcsec. We found the two produce almost identical results. While there can exist an optimal set of weights of the two factors, we stick to the closest neighbour for simplicity. Aside from detection, providing such information can also be important for the rest of the modules in the pipeline. Eventually, selecting samples depends on neighbour galaxies, given that it is conditional on measured quantities such as size and flux, affecting downstream steps such as redshift estimation. 

We trained the model twice: once with and once without incorporating the properties of the closest source. In the former case, when a source does not have a companion within $<3~\rm arcsec$, the data features corresponding to the neighbour's properties are replaced with null values. XGBoost is particularly well-suited for this scenario, as it inherently handles missing values without requiring additional preprocessing.

\begin{figure}
    \centering
    \includegraphics[width=1.\linewidth]{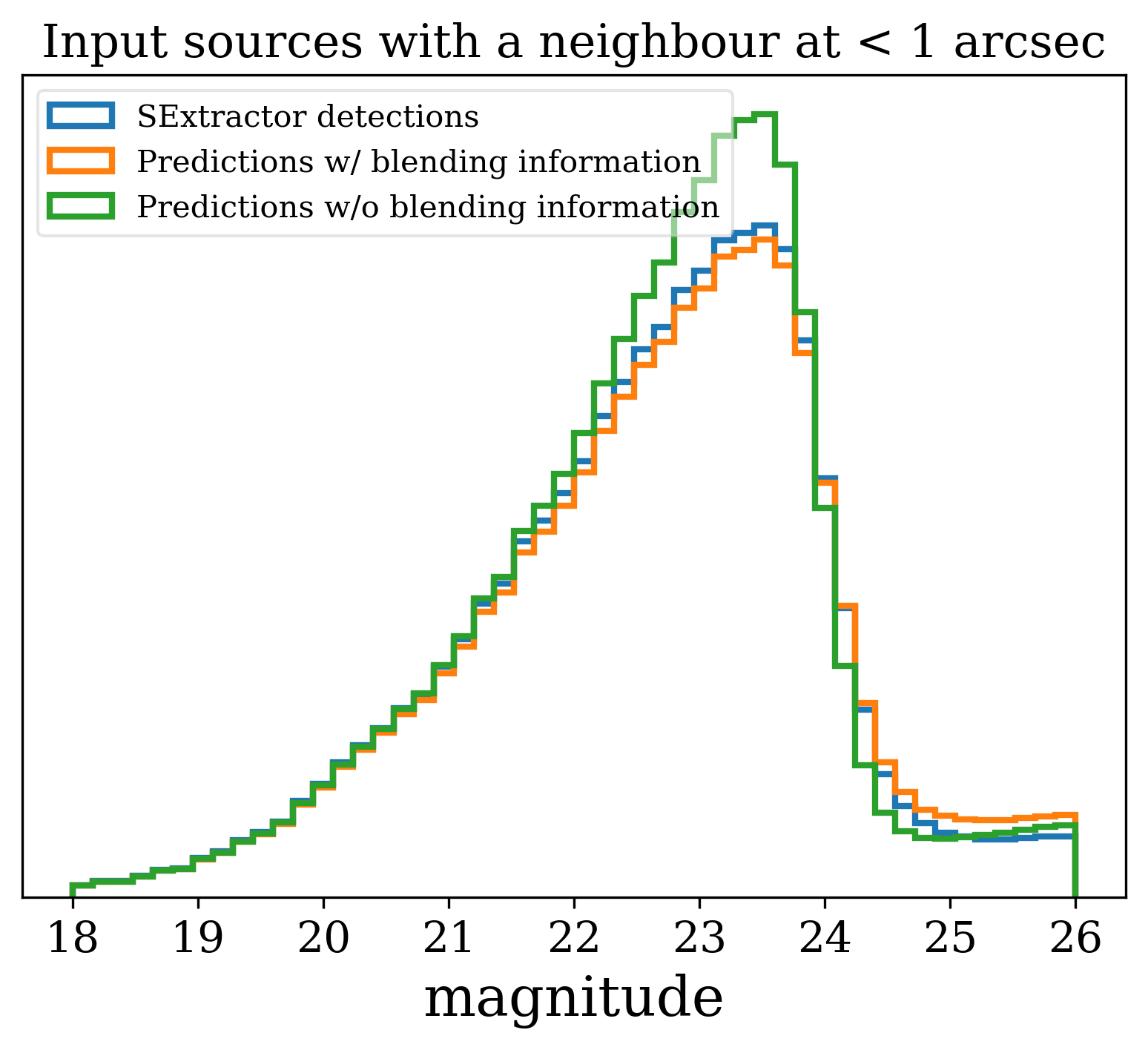}
    \includegraphics[width=1.\linewidth]{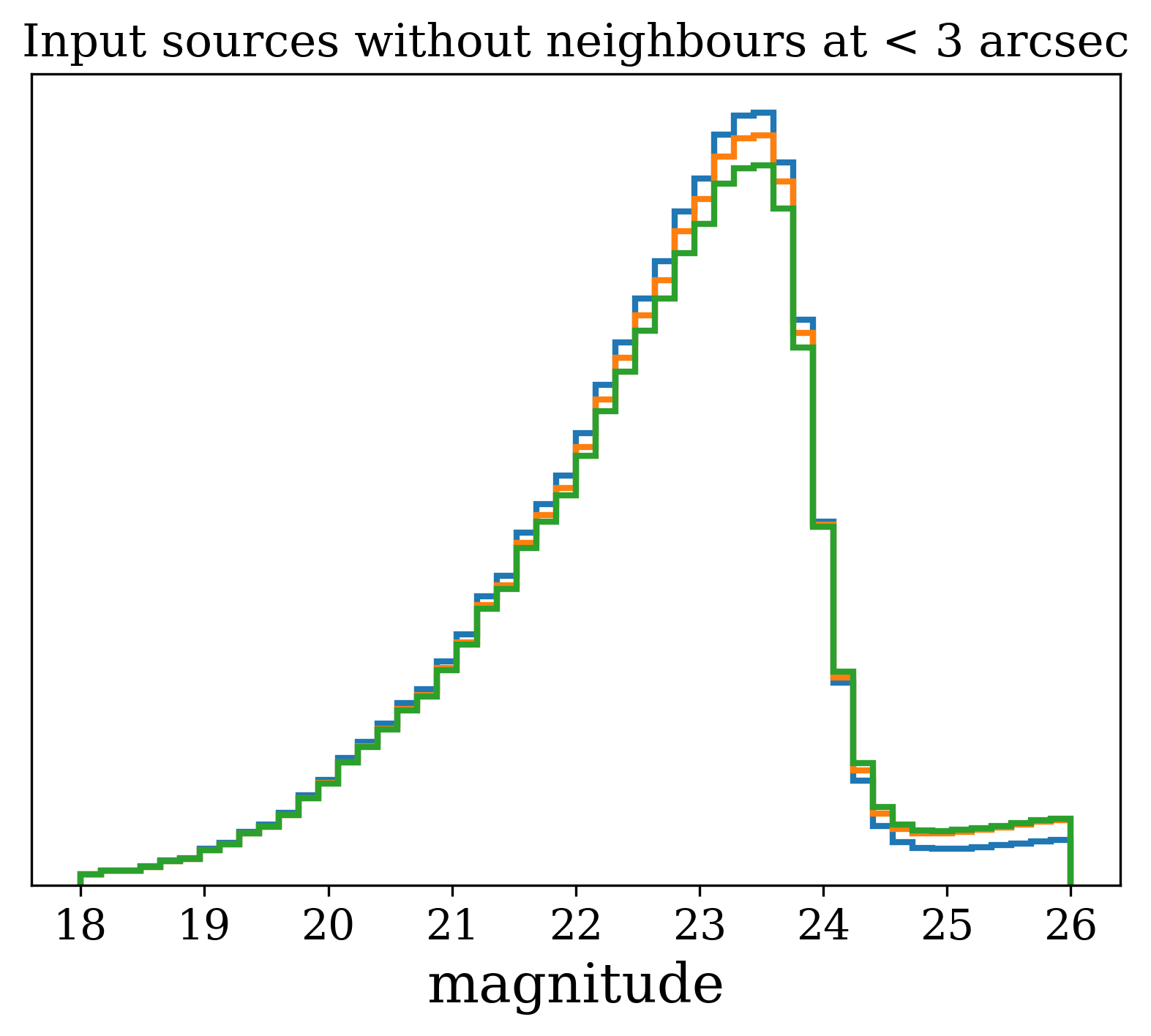}
    \caption{Histograms of input magnitude of primary galaxies associated with detections. Two groups are selected from the detected galaxies: 1) the ones with at least one neighbour found at a separation of $<1~\rm arcsec$ given their true positions (upper panel); 2) those without any neighbours found at a separation of $<3~\rm arcsec$ (lower panel). The predictions of the classifier, including blending information (orange), are shown against those of the classifier not including it (green). The results of \textsc{SExtractor} are plotted in blue as the ground truth.}
    \label{fig:detect_neighbour}

\end{figure}

To make full use of the embedded information and account for stochastic impact on detection, we used the probability output of the model instead of taking binary classification. We illustrate the importance of blending information in determining object detection in Fig. \ref{fig:detect_neighbour}. We divided the input catalogue into two groups: one consisting of sources with at least one neighbouring galaxy within $1~\rm arcsec$ and the other containing sources without any nearby galaxies within $3~\rm arcsec$. The results for the first group, obtained using two models, one trained with neighbour properties and one without, are compared in the upper panel. We observe that a significant number of sources with magnitudes between 22 and 24 are not successfully de-blended as separate objects by \textsc{SExtractor}. This effect is accurately captured by the emulator when it incorporates information about the blending environment. Interestingly, in the lower panel, the model trained with blending information still yields slightly better results for isolated sources, even though the performance of the two models appears overly similar. We further evaluate the results of the emulated detection and shear response by comparing them independently and jointly to the measurements from simulation, as  discussed in Sect. \ref{sec:mean_response} and Fig. \ref{fig:mean_R}.

\section{Results}
\label{correction}

In this section, we compare the predicted impact of blending on the effective redshift distribution between emulator and simulation. We then look into the impact of blending in different regions of galaxy parameter space as predicted by the emulator. Finally, we present our calculation of the $n_{\gamma}(z)$ values for subsets of detections grouped by their true redshifts, serving as proxies for tomographic bins.

\subsection{Mean {population-weighted} response across redshifts}
\label{sec:mean_response}

\begin{figure}
    \centering
    \includegraphics[width=\linewidth]{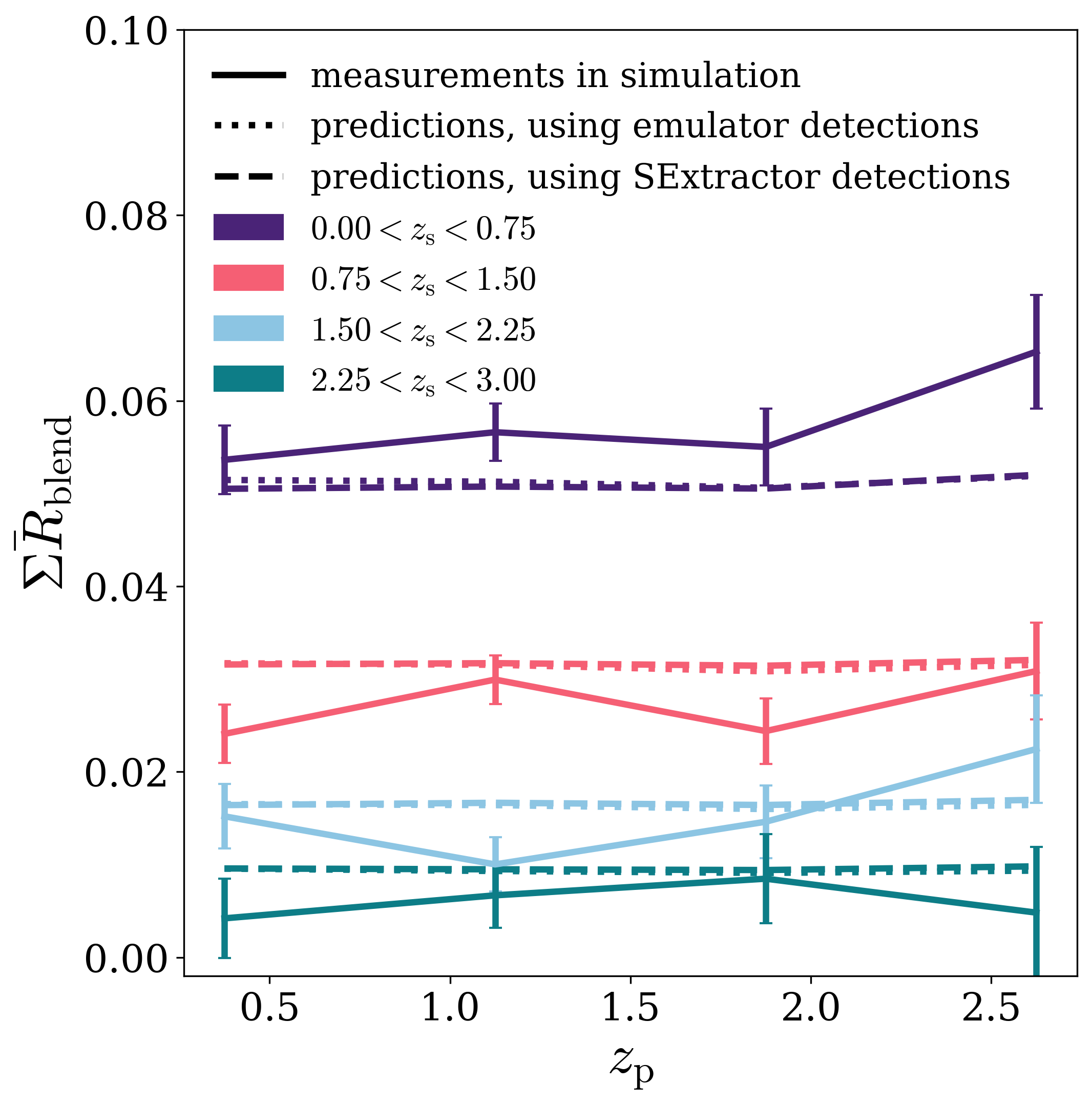}
    \caption{Averaged response of galaxy pairs as a function of the redshift, $z$, of the primary galaxy and the redshift, $z'$, of the secondary galaxy. With the measurements depicted in solid lines, the dotted ones present the results in which the responses are predictions from the emulator on the \textsc{SExtractor} catalogue, while the dashed lines use the emulator detections.}
    \label{fig:mean_R}
\end{figure}

In Fig.~\ref{fig:mean_R}, we show the response of detections to shear of secondary galaxies. Half of the input galaxies are designated as primaries, to which detections are associated. The shear responses are summed over secondary galaxies grouped into redshift intervals, and averaged over detections, which are themselves divided into the same redshift intervals. Since only half of the neighbouring galaxies are sheared and assigned as secondaries, we scaled the $\bar{R}_{\rm blend}$ in the figure by a factor of 2 relative to the measured values in our half-sheared simulation. It is also important to note that the exact values of $\bar{R}_{\rm blend}$ depend on the selection cuts applied to the sample, such as the magnitude and size, which we cut as 26 and 1.5 arcsec, respectively.

We find that $\bar{R}_{\rm blend}$ remains approximately constant with respect to the primary redshift, exhibiting only a slight positive trend. In contrast, the redshift of the secondary galaxy proves to be the more dominant factor. The averaged response increases from approximately 0.010 for secondary galaxies at $z \simeq 2$ to around 0.055 at $z \simeq 0.3$. This trend reflects our finding that the secondary galaxy dominates the impact of blending over the primary galaxy.

To validate the performance of our emulation, we compared the results using the predicted response on the \textsc{SExtractor} catalogue with those from emulator-based detections, where each input galaxy is weighted by its predicted detection probability. The two approaches show good agreement across all redshifts and are consistent with the measurements.

\subsection{Overall response over subsets of population}

\begin{figure*}
    \centering
    \includegraphics[width=0.34\linewidth]{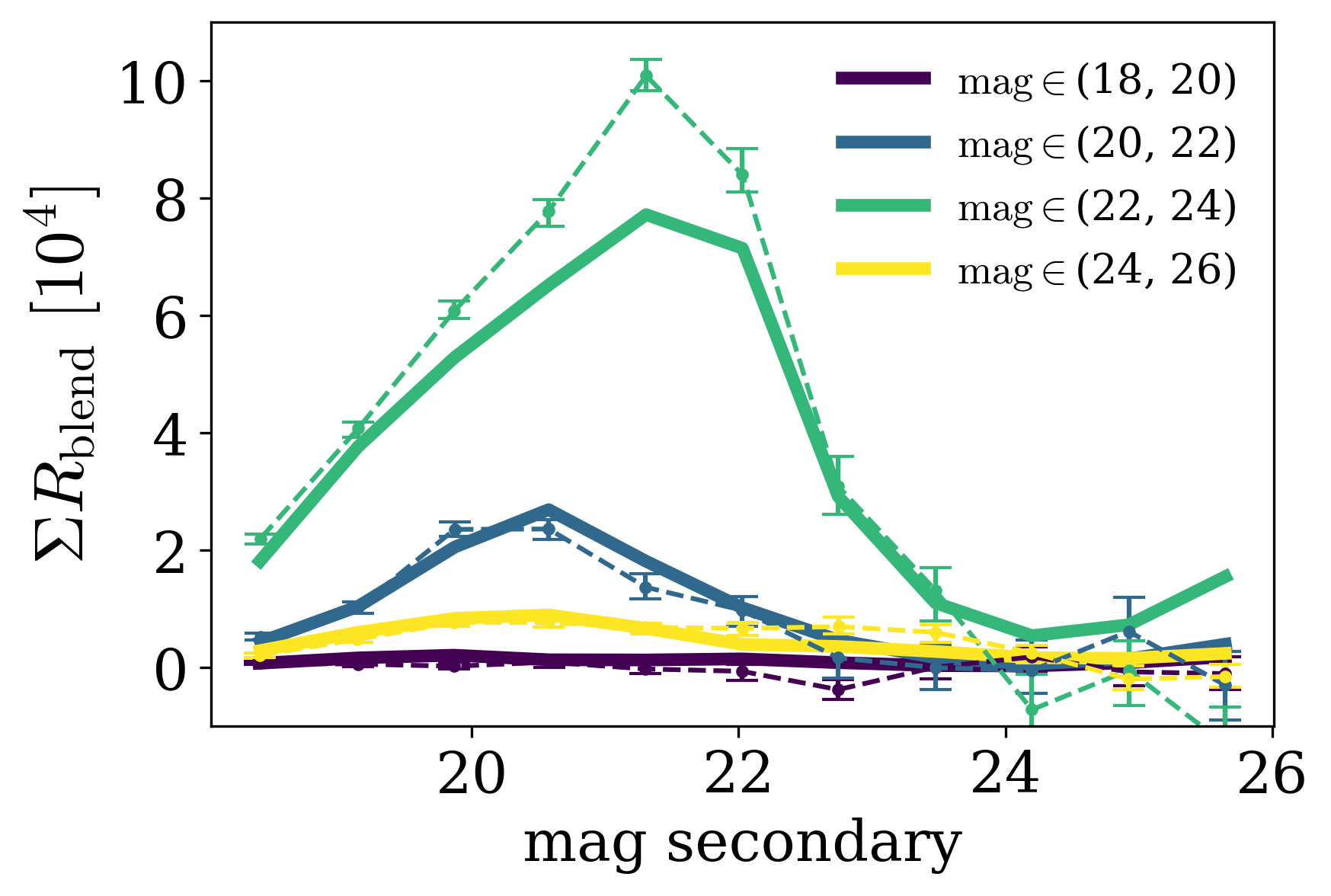}
    \includegraphics[width=0.31\linewidth]{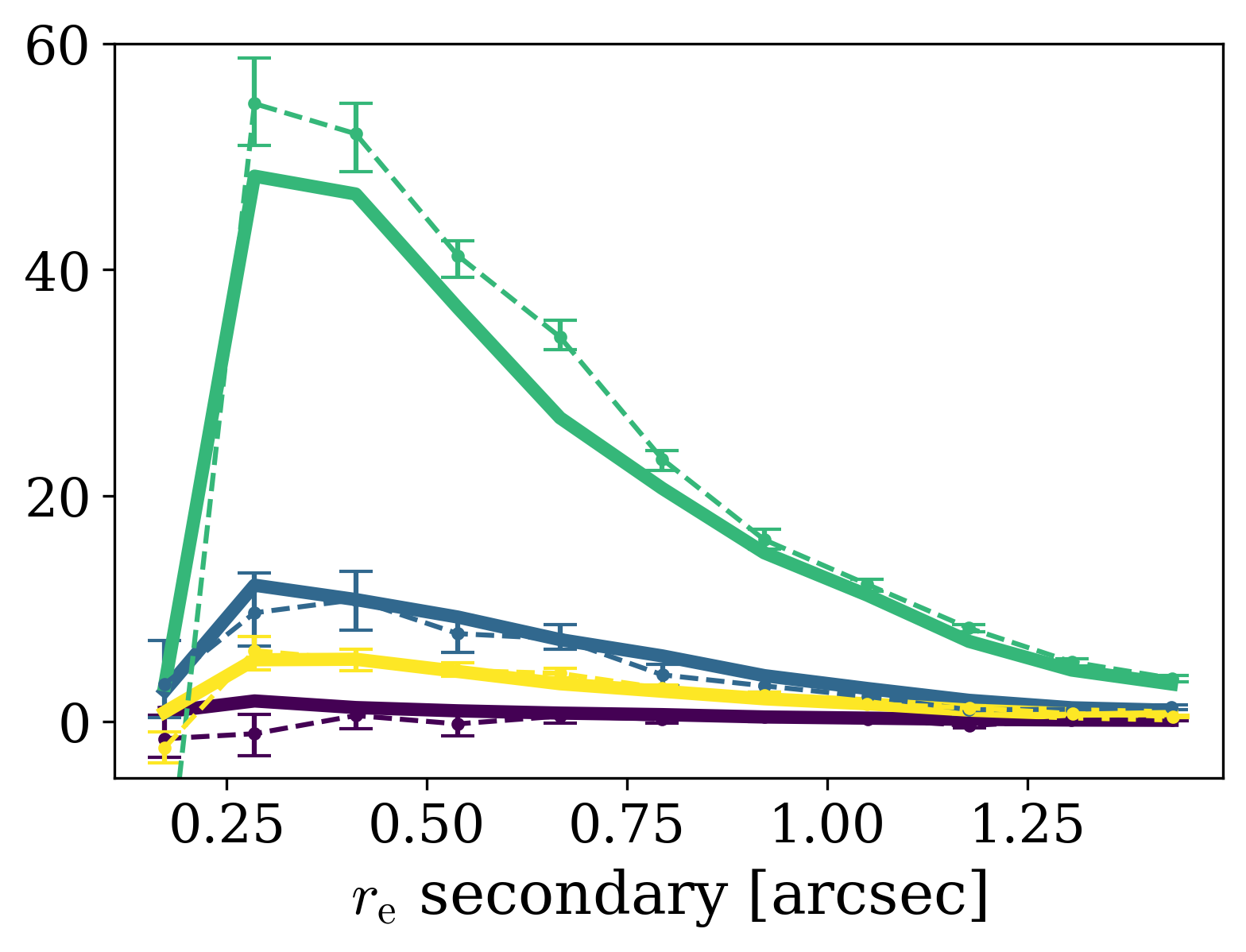}
    \includegraphics[width=0.33\linewidth]{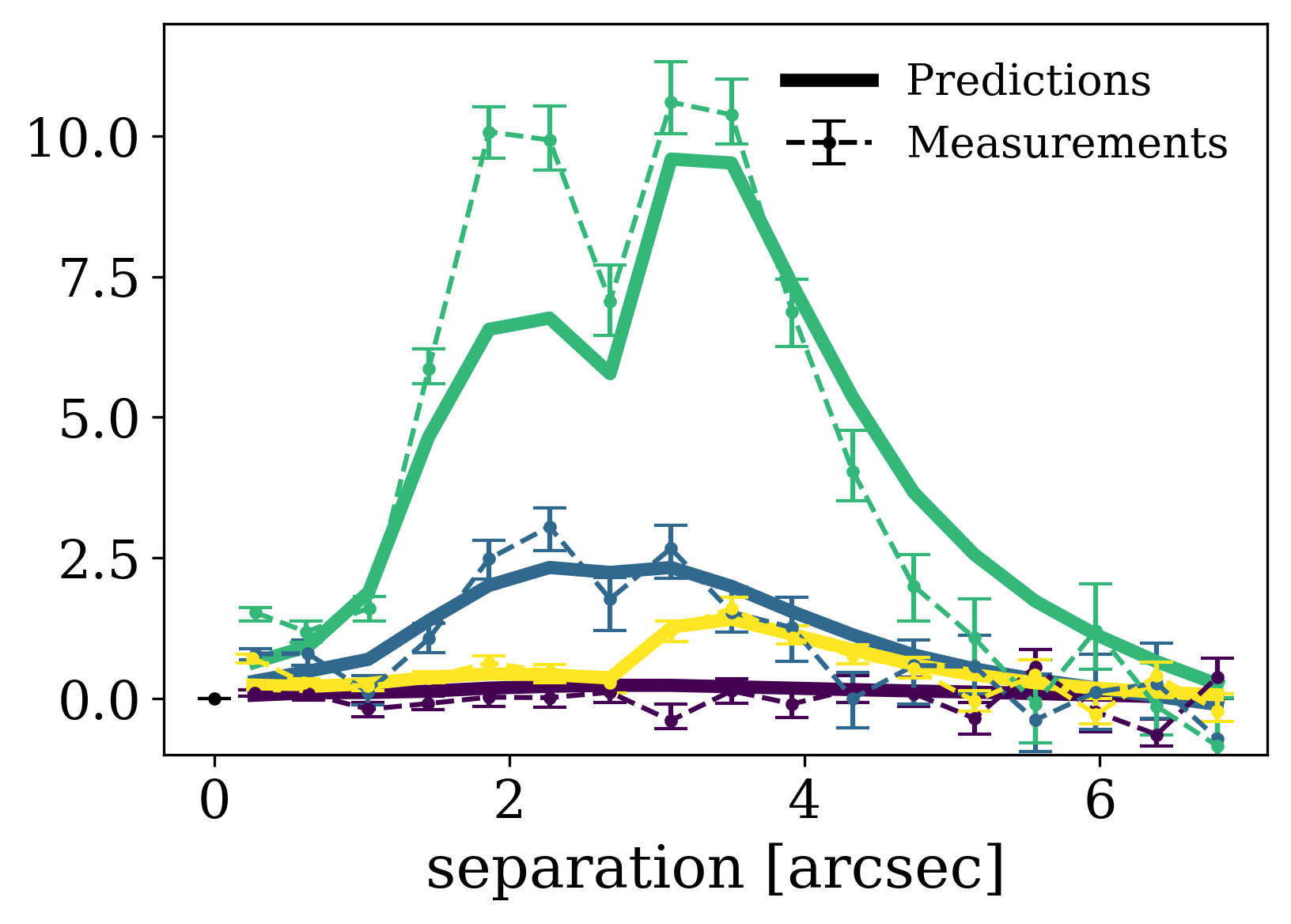}
    \caption{Emulator-predicted blending response summed over population (solid lines), in several sections with respect to the magnitude of the detected objects, as well as the magnitude, effective radius, and separation of their neighbouring galaxies. The points with error connected with dashed lines present the direct measurements from simulation.}
    \label{fig:cumu_R}
\end{figure*}

To understand where blending has the greatest impact, we also examined the summed response of detections of galaxies of different brightness over subsets of the population, in terms of the magnitude, size, and separation of secondary galaxies. In Fig. \ref{fig:cumu_R}, we present the emulator predictions versus the measurements from simulation. Both results are plotted at x-axis positions corresponding to the mean of the data within each bin.

We find that the overall effect of blending strongly depends on the brightness of the primary galaxy. The summed response increases significantly for fainter objects. It has a limited effect in the faintest magnitude bin, magnitude $\in (24, 26)$, due to incompleteness near the detection limit. Although individually they still suffer severely from blending, these detections are often removed in shear catalogues due to their low S/N. Conversely, the contribution from secondary galaxies is concentrated in the magnitude range around $\sim 21$ and size range near $\sim 0.4$ arcsec. Extremely faint or large neighbours have only a small, though non-negligible, impact. The emulator results show good agreement with the simulation measurements. Its performance slightly degrades for primary galaxies with magnitudes in the range ${\rm mag} \in (22, 24)$. There, although the emulator accurately captures the overall mean effect, the fitted model appears mildly oversmoothed, particularly with respect to secondary galaxy magnitude and pair separation. This issue could potentially be improved by training ensemble models on manually segmented datasets along these specific dimensions of the parameter space.

\begin{figure}
    \centering
    \includegraphics[width=1.\linewidth]{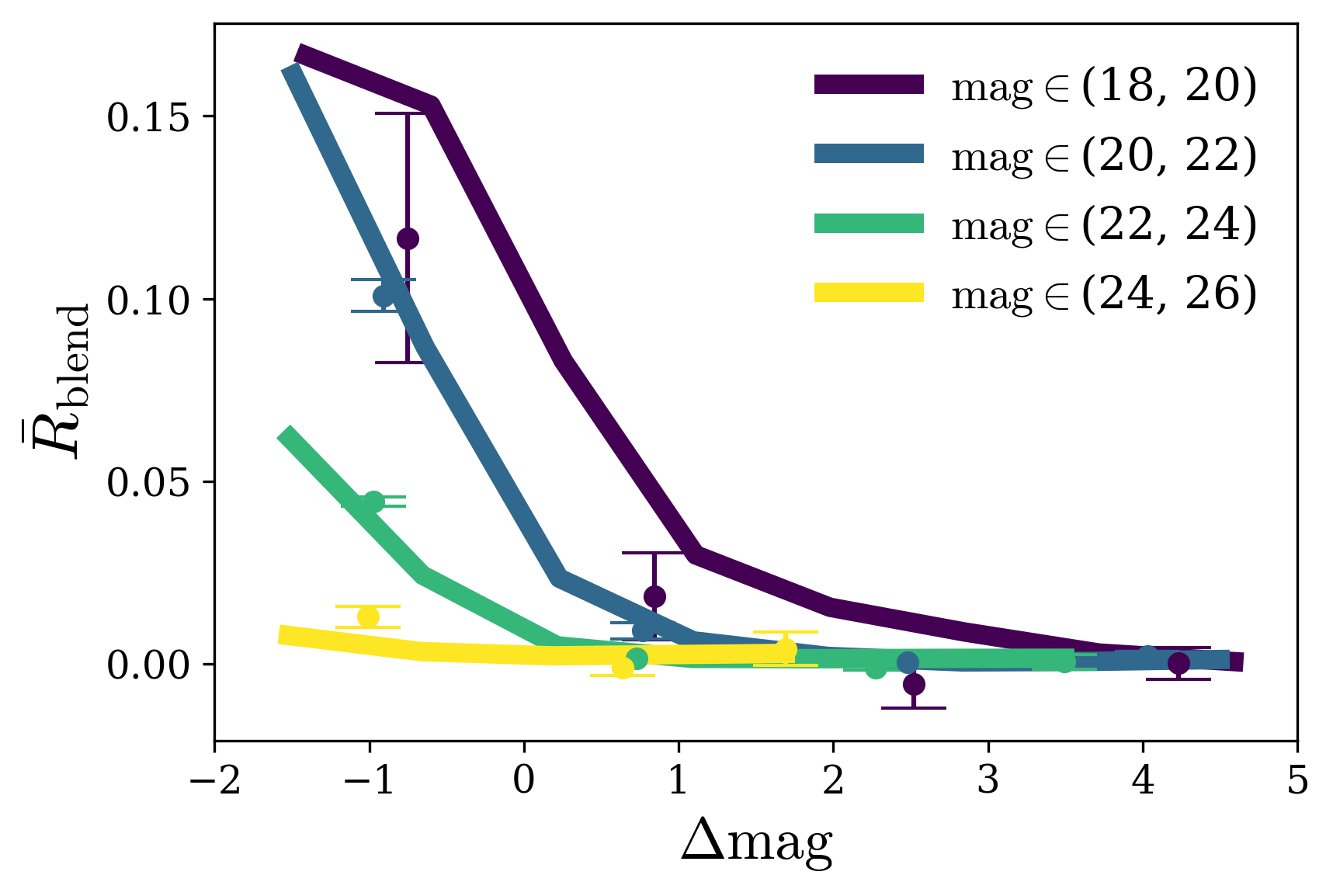}
    \caption{Mean $R_{\rm blend}$ as a function of the magnitude difference between the secondary and primary galaxies, shown for different primary magnitude bins. Curves indicate emulator predictions, including shear response and detection. Data points represent direct measurements from the simulation. Data points or parts of curves are missing where such pairs are rare.}
    \label{fig:mag_diff}
\end{figure}

Since the impact of blending is primarily driven by the relative properties of the two galaxies, we present in Fig.~\ref{fig:mag_diff} the mean response of galaxy pairs as a function of the magnitude difference, $\Delta \mathrm{mag}$, between the secondary and primary galaxies. This difference is equivalent to the logarithm of their flux ratio. We find that $\bar{R}_{\rm blend}$ drops rapidly to zero when the primary is more than one magnitude brighter than the secondary. However, the trend varies for different primary brightness bins, suggesting that $\Delta \mathrm{mag}$ alone does not fully define the effect. Since $\bar{R}_{\rm blend}$ marginalises over the distributions of other parameters, correlations between the primary galaxy's properties and its magnitude could contribute to the observed trend differences. However, they are unlikely to be the primary drivers, as $\bar{R}_{\rm blend}$ has been shown to depend only weakly on these parameters given the galaxy population (Fig.~\ref{fig:delta_e_1}).

Nearby neighbours exert the most significant influence despite their small numbers, with the response amplitude peaking at around $\sim 2$ arcsec. The ambiguous regime at separations $< 2$ arcsec remains substantial, underscoring the importance of accounting for unrecognised blends. It is important to note that the exact values of these limits are survey-specific and dependent on observing conditions.

\subsection{$n_{\gamma}$ correction}

\begin{figure*}
    \centering
    \includegraphics[width=\linewidth]{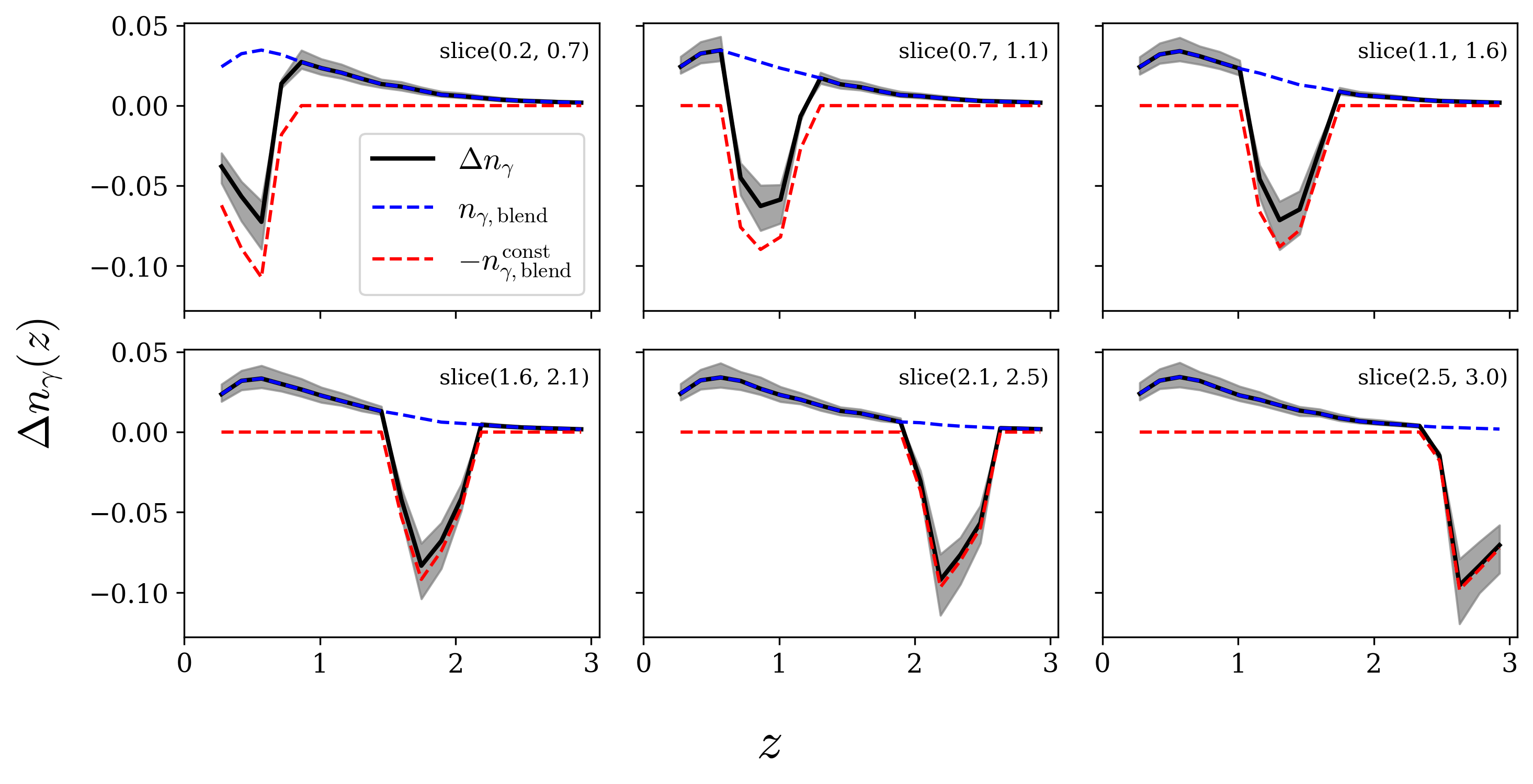}
    \caption{Modelling $\Delta n_{\gamma}(z)$ on  true galaxy catalogue using our emulator. The detections are divided into six slices based on their true redshifts. The black line shows our correction (Eq. \ref{eq:correction}) of each slice on shear calibration that ignores redshift dependence. The grey area around the black line presents the 1-$\sigma$ uncertainty of the galaxy population model. The correction decomposes into the effect of secondary galaxies $n_{\gamma,{\rm blend}}$ (dashed blue curves) and the falsely adopted self-response $n_{\gamma,{\rm blend}}^{\rm const}$ from `constant-shear' calibration (dashed red curves).}
    \label{fig:n_gamma}
\end{figure*}

In Fig. \ref{fig:n_gamma}, we compute $n_{\gamma, \rm blend}$ as the contribution of blending to the effective redshift distribution, and $\Delta n_{\gamma, \rm blend}$ defined in Eq. \ref{eq:correction}, as the correction on traditional calibrations where the effect of varying shear on component galaxies is ignored. In this work, we use the true redshifts of galaxies and group them into six uniform intervals within the range $[0.2, 3.0]$. In real observations, where galaxies are binned based on photometrically estimated redshifts, the resulting distributions with respect to true redshifts are typically noisy and broadened, which tends to dilute the features observed in our analysis.

We observe a dip in $\Delta n_{\gamma, \rm blend}$ that follows the redshift of the samples, along with slice-independent positive values outside the sample interval. The total integral remains zero by construction of the formulation. {This behaviour is expected from the definition of $\Delta n_{\gamma, \rm blend}$ in Eq. \ref{eq:correction}} as the difference between two cases: $n_{\gamma, \rm blend}^{\rm}$, where the blending response is correctly assigned to the lens (secondary) redshift, and $n_{\gamma, \rm blend}^{\rm const}$, where it is incorrectly attributed to the sample (primary) redshift. The latter represents the outcome expected from a `constant-shear' calibration or algorithms that do not separate $R_{\rm self}$ and $R_{\rm blend}$. In such cases, the `local' signal at the primary galaxy's redshift is overestimated by the product of $R_{\rm blend}$ and the primary shear, manifesting as dips in the correction. {A direct numerical comparison to the findings in Fig. 10 of \cite{maccrannY3ResultsBlending2022} is not possible, since their response is evaluated after full photo-z binning and survey-level selections. Nevertheless, we find qualitatively consistent trends: negative `local' corrections on the \textsc{Metacalibration} $n_{\gamma}(z)$ of each bin and the overall magnitude of the effect is of the same order.}

In contrast, $n_{\gamma, \rm blend}^{\rm}$ reflects the contribution of responses weighted by the shear at the secondary redshift, determined by the secondary galaxy redshift distribution of blends in each bin. Based on conclusions from previous sections that $R_{\rm blend}$ primarily depends on secondary galaxies across the population, $n_{\gamma, \rm blend}^{\rm var}$ remains largely constant across sample slices.

It is straightforward to calculate the following quantity that is related to the shift in mean effective redshift for each slice, $i$,
\begin{equation}
    \Delta \bar{z}_i = \int dz~z~\Delta n_{\gamma, i}(z).
\end{equation}
We note that $\int dz~z~n_{\gamma}(z)$ represents only a pseudo-mean effective redshift that approximates the true effective $\bar{z}$. This is because, in this work, we normalised $n(z)$ rather than $n_{\gamma}(z)$, due to the limitation that we do not model $R_{\rm self}$ and therefore lack knowledge of the absolute value of $n_{\gamma}$. As a result, the above expression does not strictly quantify the shift in the effective mean redshift. Since the total shear response, $R$, is typically less than 1, this estimate of $\Delta \bar{z}$ is expected to underestimate the actual shift in the mean effective redshift distribution. Furthermore, we neglect the change in the normalisation of $n_{\gamma}(z)$, which could, in principle, be altered by shear-induced detection effects in blends when considering varying shear of the galaxies. However, we expect such effects to be minor, and thus the normalisation of $n(z)$ remains effectively unchanged relative to the constant-shear case.

We present the change in the approximated mean effective redshift across redshift slices in Table  \ref{tab:mean_z}. A roughly linear relationship is observed between $\Delta z$ and the sample redshift: the shift decreases from $0.023_{+0.006}^{-0.004}$ in the $(0.2, 0.7)$ bin to $-0.055^{+0.010}_{-0.012}$ in the $(2.5, 3.0)$ bin. The whole sample sees a change of $-0.009^{+0.002}_{-0.001}$. The error bars reflect population uncertainty, estimated by recomputing the results using catalogues based on the SMF16 and SMF84 models, while shot noise is negligible and excluded from the error estimation. These values present a tension with the current accuracy of redshift characterisation (e.g. \citealt{TanakaHSCRedshift2018,HildebrandtKiDSRedshift2021,MylesDESredshift2021}). The uncertainty arising solely from galaxy population variations significantly exceeds the requirements of mean redshift bias for next-generation surveys, for instance, $0.001(1+z)$ for LSST Year 10 \citep{LSSTSRD2018}, highlighting the necessity of accounting for this effect in future evaluations.

It is important to note that, since we use true redshifts, photo-$z$ characterisation and binning will significantly affect the final values. Binning primarily acts to dilute the true redshift slices; therefore, broader bins can smear the $\Delta \bar{z}$ values across slices, making them appear more uniform. Additionally, factors such as how blends are treated, whether omitted or de-blended when recognised, can affect the amplitude of the shift. Nonetheless, these redshift shifts and the uncertainty on these shifts caused by uncertainty in the galaxy population represent a significant source of systematic error, posing an urgent concern for final stage-III analysis and upcoming stage-IV surveys.

\begin{table}
  \centering
  \caption{Shifts in {approximated} mean effective redshift for six slices of detected galaxies with our $\Delta n_{\gamma}$ model.}
  \label{tab:mean_z}
  \begin{tabular}{c c c c}
    \toprule
    slice & {$\Delta \bar{z}$} & input (\%) & detected (\%) \\
    \midrule
    \rule{0pt}{4ex}(0.20, 0.67) & \hspace{0.6em}$0.023^{-0.004}_{+0.006}$ & $18.18^{-0.04}_{+0.10}$ & $7.76^{+0.82}_{-0.72}$ \\
    \rule{0pt}{4ex}(0.67, 1.13) & \hspace{0.6em}$0.004^{-0.001}_{+0.001}$ & $25.44^{+0.06}_{-0.18}$ & $9.00^{+1.26}_{-1.08}$ \\
    \rule{0pt}{4ex}(1.13, 1.60) & \hspace{0.0em}$-0.015^{+0.002}_{-0.003}$ & $22.28_{-0.22}^{+0.13}$ & $7.27^{+0.99}_{-0.89}$ \\
    \rule{0pt}{4ex}(1.60, 2.07) & \hspace{0.0em}$-0.032^{+0.006}_{-0.007}$ & $15.54^{-0.02}_{+0.07}$ & $4.69^{+0.70}_{-0.56}$ \\
    \rule{0pt}{4ex}(2.07, 2.53) & \hspace{0.0em}$-0.052^{+0.009}_{-0.012}$ & $10.75^{+0.14}_{-0.01}$ & $2.83^{+0.48}_{-0.36}$ \\
    \rule{0pt}{4ex}(2.53, 3.00) & \hspace{0.0em}$-0.055^{+0.009}_{-0.013}$ & $7.81^{+0.08}_{-0.12}$  & $1.87^{+0.26}_{-0.25}$ \\
    \bottomrule
  \end{tabular}
  \tablefoot{The upper bound (lower bound) of errors refers to galaxy population model SMF16 (SMF84), while shot noise is excluded. The fractions of input galaxies and detected galaxies after applying both the size and magnitude cuts in each slice, relative to the full input catalogue, are also shown.}
\end{table}

\section{Discussion}

{In this work we introduce a pair-based framework to emulate the blending shear response and to estimate the resulting shift in the effective redshift distribution. While the method is built on solid ground, it is useful to summarise the main assumptions and comment on their limitations.}

{1. Linear pairwise approximation:
We assume that the response of a blended system can be decomposed into a sum of pairwise contributions. As explained in Appendix~\ref{sec:linearity}, this approximation weakens in rare cases involving multiple bright neighbours. Our simulations naturally capture the average impact of such events at the depth and density of our setup. Still, significantly deeper or denser surveys may require additional validation or a revised emulator trained on those conditions.}

{2. Simplified PSF and noise model.
Our simulations adopt a spatially constant PSF and fixed noise level, whereas real surveys exhibit spatial PSF variation, correlated noise, and a non-trivial background estimation. Appendix~\ref{sec:invariance} shows that rescaled galaxy properties remain consistent in terms of the blending response and detection probability, indicating that our emulator can generalise under moderate variability of observing conditions. Higher order effects from PSF modelling residuals may become relevant for higher precision, although Appendix~\ref{sec:PSF} demonstrates that $R_{\rm blend}$ is largely insensitive to PSF ellipticity. Addressing these effects does not require retraining a new emulator.}

{3. Neglect of magnification and foreground-background alignment.
Our emulator operates on catalogue-level mocks, where, in this work, we use unmagnified galaxy properties and uncorrelated positions. This neglects magnification-induced changes in number density and flux, as well as alignments between structures at different redshifts. Both effects can modify the abundance and influence of secondary galaxies and therefore alter the cumulative $R_{\rm blend}$. These effects are spatially coherent and may interact with physically driven detection bias. Extending the method to include magnification-aware neighbour modelling and correlated large-scale-structure mocks represents a natural future improvement. Addressing such effects does not require retraining a new emulator.}

{4. Application to cosmological analyses.
With realistic binning and selection, our method can be readily applied to current datasets. However, for surveys with substantially increased depth, new simulations may be required to avoid extrapolating beyond the galaxies represented in the current training data. Furthermore, the impact of blending depends on the survey’s specific image-processing pipeline. Pixel-level masking, flux assignment, de-blending choices, and the shear measurement method all modify how neighbour light enters detection and shear inference. A faithful cosmological application therefore requires applying our correction within a survey-consistent simulation and processing framework, ensuring that the detection, measurement, and binning steps are handled coherently.}

\section{Conclusions}

In this work, we propose a new approach to addressing the problem of redshift-mixing blending where the weak-lensing signal leaks across redshift. The effective redshift distribution is no longer solely dependent on the redshift of the primary galaxy detected, but also on the redshift of neighbouring galaxies. Instead of generating full suites of realistic simulations with a redshift-varying shear field, we decomposed the problem to the level of individual galaxy pairs. This offers flexibility in terms of accounting for the uncertainty of our galaxy population model in the calibration and allows us to correct the effective redshift distribution in fine-grid redshift steps without assuming a parametric model.

We measured the blending response, $R_{\rm blend}$, of galaxy pairs with various possible configurations in simulations, where half of the input galaxies are designated as secondaries and sheared in random directions. Then $R_{\rm blend}$ was calculated as the change in the shapes of detections associated with the primary galaxies in response to the applied shear of the secondary galaxies. We emulated the measurements for each pair of secondary and primary galaxies using gradient boosted trees. To generalise the model under varying observing conditions, both within and beyond a survey, we rescaled the galaxies’ properties with respect to background noise and PSF size. We verified that $R_{\rm blend}$ remains invariant under this transformation. This allows for our model to be easily extended to studying the sensitivity of blending to observing conditions.

We trained the model on a 200-square-degree simulation and find that its predictions are consistent with the direct measurements within $1\sigma$. In addition to accounting for population uncertainty in calibration, we must also address the domain shift problem, where the training population may be distributed differently compared to real data. To tackle this, we simulated additional sets using $1\sigma$ quantiles of a population model calibrated from observations, whereas the training data use the median value. We found the average $R_{\rm blend}$ drops by $\sim39\%$ from the higher $1\sigma$-quantile to the lower $1\sigma$-quantile. This is mostly due to the variations in the faint end of the stellar mass function. Hence, uncertainties in galaxy evolution exert a substantial impact on shear calibration. We demonstrated that our $R_{\rm blend}$ emulator has sufficient sensitivity to capture such variations.

Aside from shear response, the full implementation of our method requires emulating the galaxy selection function, including object detection, selection cuts, and redshift binning. Due to the complexity introduced by the correlation between measured properties and blending, it is more straightforward to build the emulators based on the true quantities. We began with the detection step, finding that incorporating information from neighbouring objects significantly improves performance. This motivates the use of such information in emulating various downstream processes that rely on measured properties affected by blending.

Finally, by examining the averaged responses across source and lens redshifts, we found that $\bar{R}_{\rm blend}$ strongly depends on the secondary redshift, decreasing from approximately 0.03 at $z=0.3$ to about $0.005$ at $z=2.0$, while exhibiting a negligible trend with the primary redshift. The exact values depend on the applied cuts and the sample population. This behaviour propagates into our calculation of the $n_{\gamma}(z)$ correction. We find it to consist of two components: one dependent on and the other independent of the sample redshift distribution. The former compensates for the overestimation of `local' signals in each bin introduced by `constant-shear' calibration. The latter is the contribution of blended secondary galaxies predicted by our model, which remains nearly constant between bins in the primary galaxy's redshift. We calculated the shifts in the approximated mean effective redshift within each primary galaxy redshift bin and found $\Delta \bar{z}$ to range from approximately $0.02$ to $-0.05$. The exact forms of $\Delta n_{\gamma}(z)$ and $\Delta \bar{z}$ depend on the specific characterisation of each tomographic bin in actual surveys, which we postpone to works in the near future. Beyond that, it is also worthwhile to estimate the relative contribution of this effect, given our model, to the biased cosmological inference in Stage-IV surveys.

Taken together, our findings affirm the large impact of blending. They indicate that emulation is part of a viable method of accounting for this effect in interpreting measurements, especially as galaxy surveys grow in depth and volume. In weak lensing studies, where correlations of observed galaxy shapes across redshift bins are measured, blending can impact every step, from shape measurement to sample selection. Beyond the redshift-dependent shape bias explored in this work, properties such as flux and size (essential for sample cuts and binning) can also be modelled with similar techniques. Although these effects are often treated as spatially uniform by default, spatial variation in blending should be accounted for alongside the consideration of angular dependencies. This also raises interest in possible correlations between blending and the matter density field, as blending may introduce stronger detection biases in overdense regions \citep{GencBlendDetect2025}. Our emulator, which integrates information from neighbouring galaxies on an individual basis, provides a practical tool for such analyses.

\begin{acknowledgements}

This work was supported by the Excellence Cluster ORIGINS, funded by the Deutsche Forschungsgemeinschaft (DFG, German Research Foundation) under Germany’s Excellence Strategy – EXC-2094 – 390783311. ZZ acknowledges support from the German Academic Exchange Service (DAAD) for this project. SSL acknowledges funding from the programme ``Netzwerke 2021'', an initiative of the Ministry of Culture and Science of the State of Northrhine Westphalia and support from the European Research Council (ERC) under the European Union’s Horizon 2020 research and innovation program with Grant agreement No. 101053992.

\end{acknowledgements}

\section*{Data availability}

The data used in this study will be shared by the corresponding author upon reasonable request. The emulator is available at \url{https://github.com/zhangzzk/blending_emulator}.


\bibliographystyle{aa}
\bibliography{references}



\appendix

\section{Linearity of secondary shear response}
\label{sec:linearity}

\begin{figure}
    \centering
    \includegraphics[width=1.\linewidth]{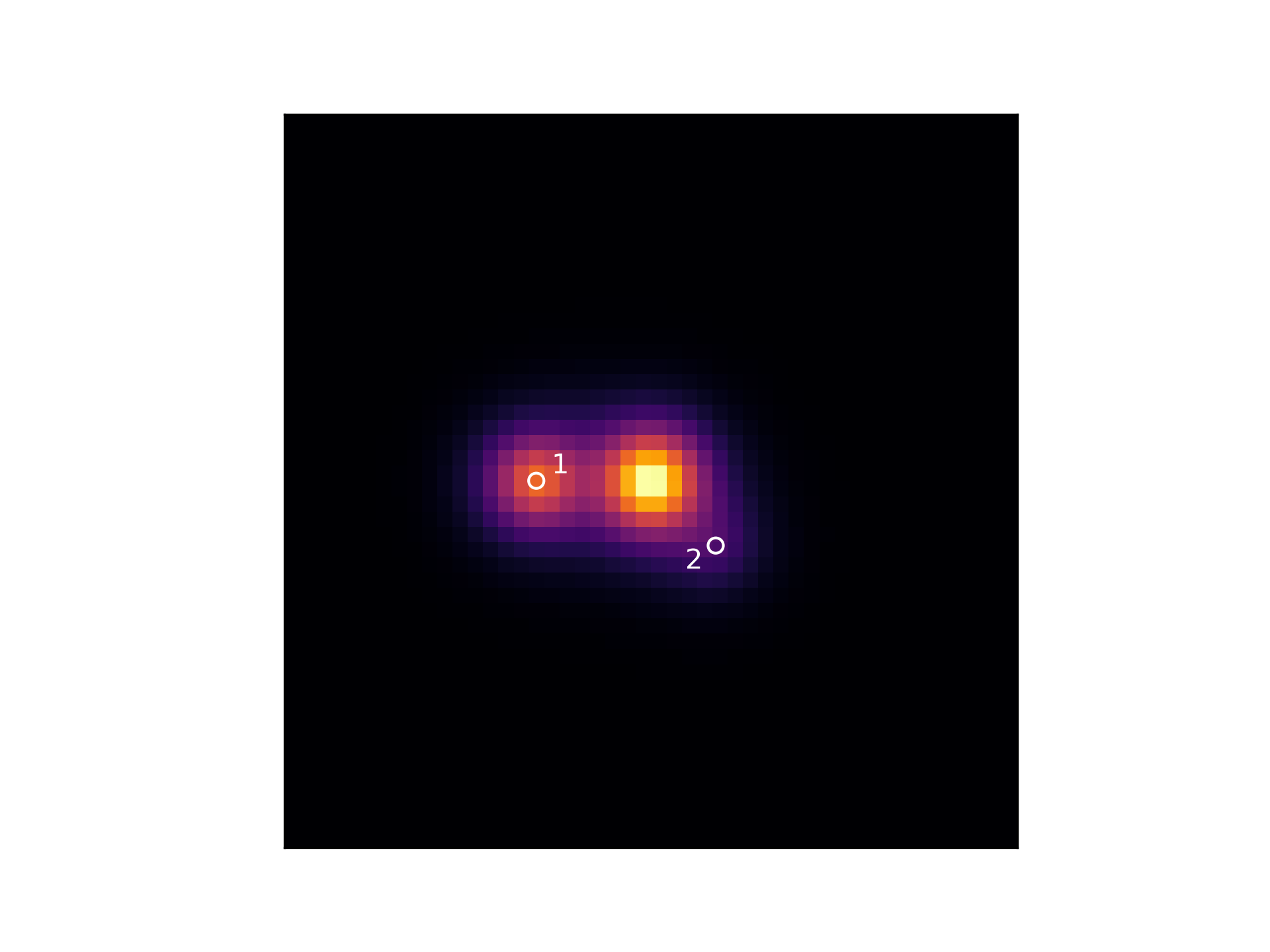}
    \caption{Image of a simulated triplet blending on a postage stamp. The primary galaxy has two close companion neighbours, marked with a small white circle and an index for each.}
    \label{fig:linearity_example}
\end{figure}

The way we write down the effective redshift distribution as in Eq. \ref{eq:g_blending_z}, integrating the blending effect over the secondary galaxy population includes two implicit assumptions on the linearity of $R_{\rm blend}$: 1) The measured ellipticity linearly scales with the shear applied to the secondary galaxy, 
\begin{equation}
\label{eq:assumption1}
    \delta \textbf{e} = \textbf{R} \times \textbf{g}_{\rm secondary}.     
\end{equation}
2) When more than one secondary galaxy is present, the measured ellipticity responds to the linear combination of the separate contributions from each secondary galaxy, 
\begin{equation}
\label{eq:assumption2}
    \delta \textbf{e} = \sum_i \textbf{R}_i \times \textbf{g}_{i{\rm th}~{\rm secondary}}.
\end{equation}
We validated these assumptions in simplified toy simulations where galaxies are drawn on postage stamps, as shown in Fig. \ref{fig:linearity_example}. We draw three galaxies, the primary, a first neighbour and a second neighbour, respectively with magnitudes 20, 20.5, and 22. They shear as $r_e=0.4$ arcsec, Sersic index 1, and BA=1. The separation of the two neighbours from the primary is 1.5 and 1.2 arcsec, respectively. The PSF used is the same model as in the main simulation. The background noise is set to zero for simplicity. 

\begin{figure*}
    \centering
    \includegraphics[width=0.48\linewidth]{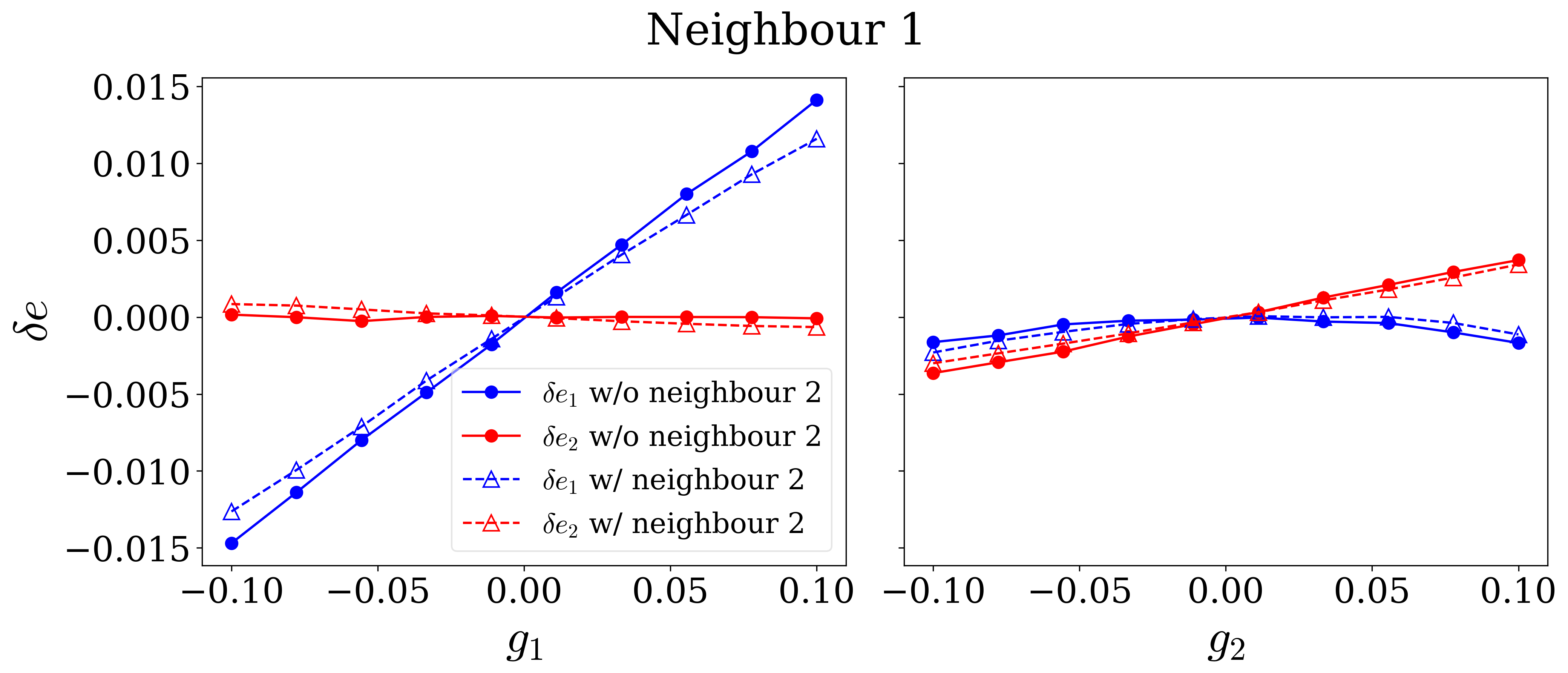}
    \includegraphics[width=0.48\linewidth]{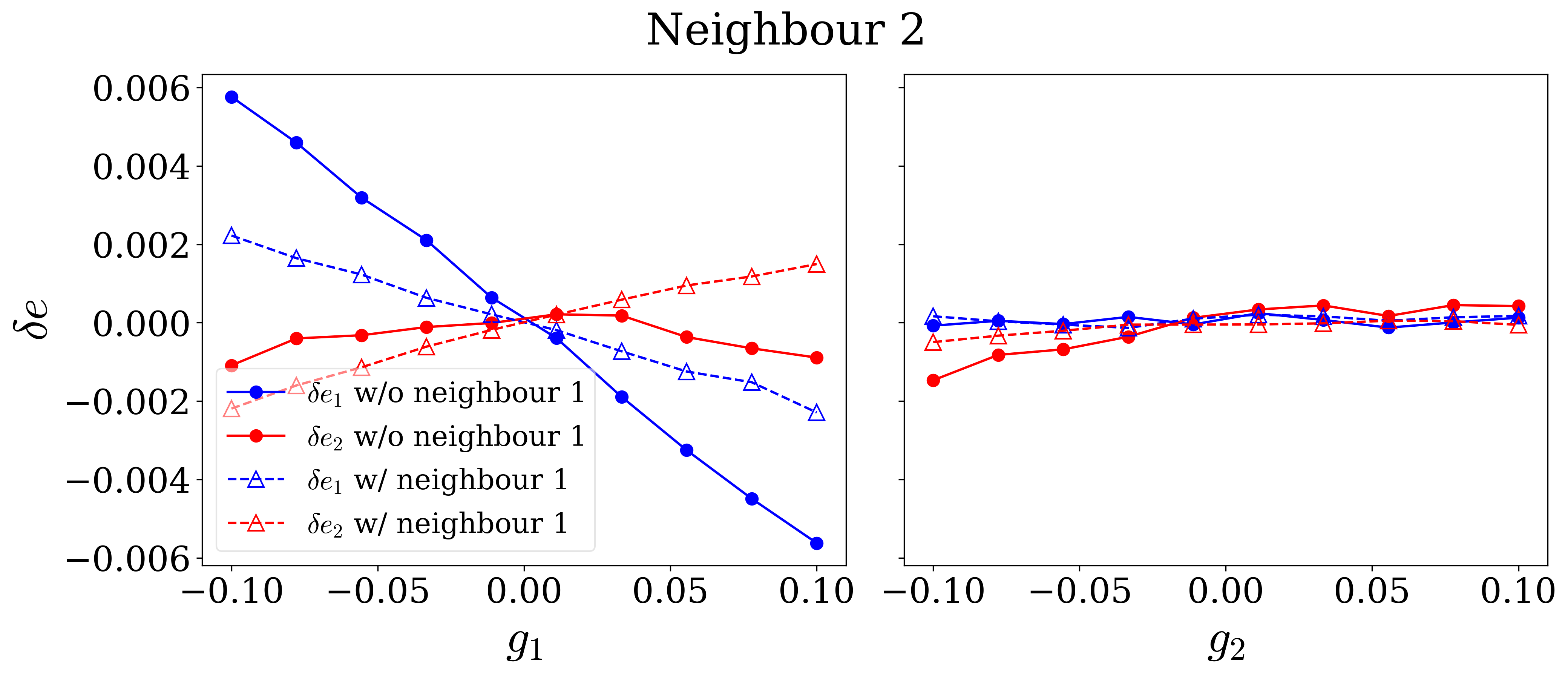}
    \caption{Tests on the linearity of the measured galaxy shape against the shear applied on the first neighbour (upper panel) and the second neighbour (lower panel). Solid lines connecting the dots are the results when the third galaxy is removed, while the dashed lines connecting the hollow triangles are when it is present. Blue points are the first component of the galaxy ellipticity, and red points are the second. The elements of $\textbf{R}$ to each neighbour are calculated by fitting the points to a linear function.} 
    \label{fig:linearity_test1}
\end{figure*}

\begin{figure}
    \centering
    \includegraphics[width=0.9\linewidth]{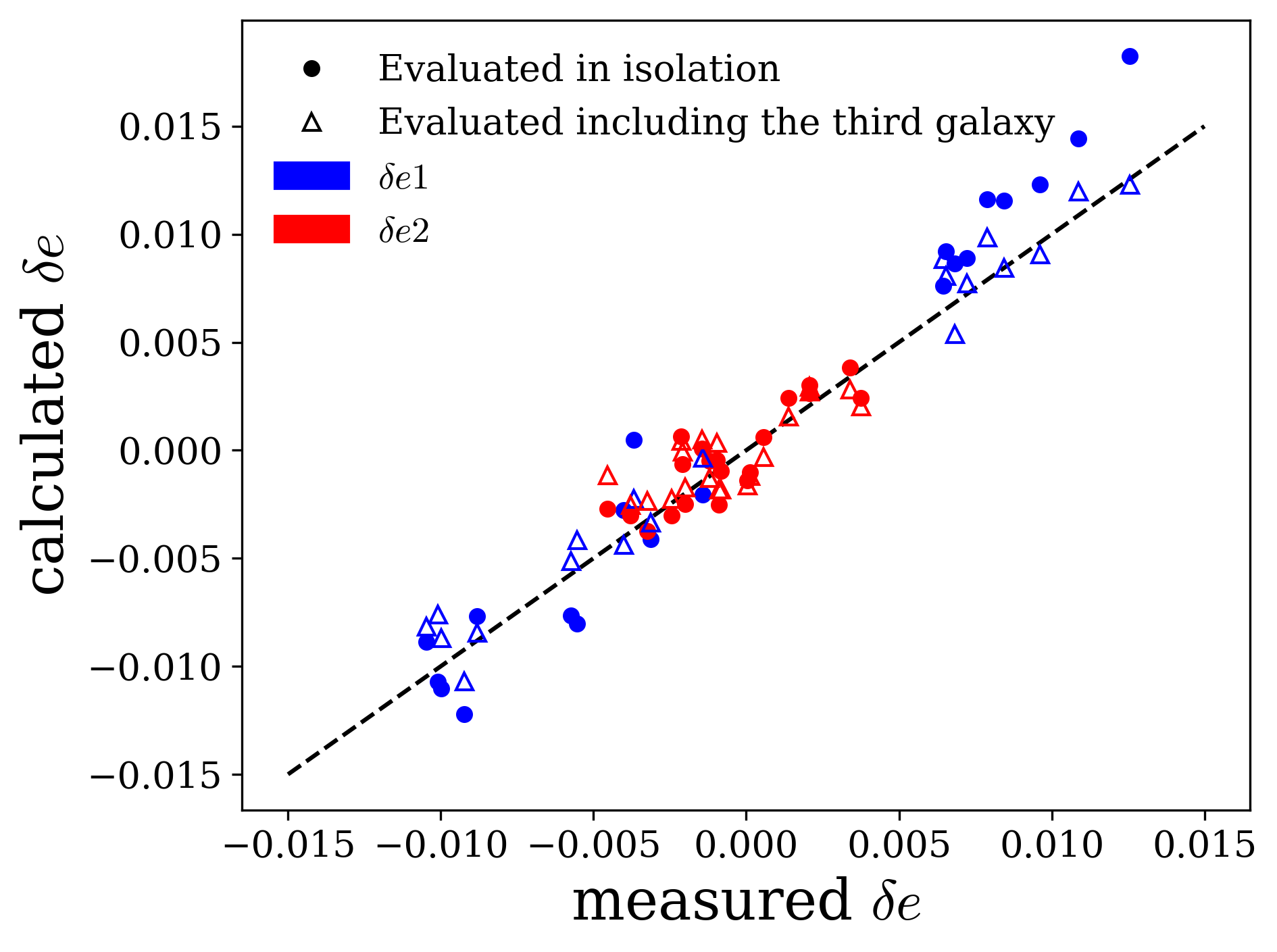}
    \caption{Calculated difference in the measured galaxy shape when randomly shearing both secondary galaxies against the measured results. Blue points are the first component of the galaxy ellipticity, and red points are the second. Solid dots show the results where the calculations combine the evaluated $\textbf{R}$ of each neighbour, removing the third galaxy. Hollow triangles are when using $\textbf{R}$ with the third galaxy turned on.}
    \label{fig:linearity_test2}
\end{figure}

We first calculated the isolated $\textbf{R}_i$ of each secondary galaxy by shearing it with different values from -0.1 to 0.1 independently in two components, removing the other secondary. The results are shown as solid lines in Fig. \ref{fig:linearity_test1}. $\delta e$ of both components in both cases are observed to have a linear relation with the applied shear, while noticeable quadratic behaviours are also present in cases where the slope is small. $\textbf{R}_i$ for each neighbour is derived by finding the slope of the points. To test the second assumption, we then shear the two neighbours simultaneously with random amounts and directions and measure $\delta e$, which is compared with the calculated results using Eq. \ref{eq:assumption2} and the pre-derived $\textbf{R}_i$. The results are shown as filled points in Fig. \ref{fig:linearity_test2}. The calculations exhibit a biased linear relation to the measurements, indicating that the isolated evaluation of $\textbf{R}_i$ may be reduced by the light of the third galaxy. In extreme cases, such as when both secondary galaxies are significantly brighter than the primary, the bias can become substantial.

To confirm this, we remeasured $\delta e$ as shown in dashed lines in Fig. \ref{fig:linearity_test1} by shearing each secondary without removing the other. The linearity is retained in the new tests, while the slopes show declines in most cases, especially when the third galaxy is bright. We then repeat the fitting of the new slopes and compare the calculations of $\delta e$ using the newly evaluated $\textbf{R}_i$, considering the presence of the third galaxy. As shown as hollow triangles in Fig. \ref{fig:linearity_test2}, the calculated results align well with the measurements. This test suggests that our second assumption does not necessarily hold when the primary galaxy is neighbouring bright companions. However, due to its rarity, this violation would have minimal impact on the overall effect. Furthermore, the way we simulate $\textbf{R}_i$ by shearing galaxies on a realistic sky already accounts for the impact of the response of the possible occurrence of a third or additional galaxies for each primary and secondary pair. Given that the number of bright galaxies remains relatively consistent across different population models, our emulator should capture this effect with reliable accuracy.

\section{Sensitivity to a PSF}
\label{sec:PSF}

\begin{figure}
    \centering
    \includegraphics[width=1.\linewidth]{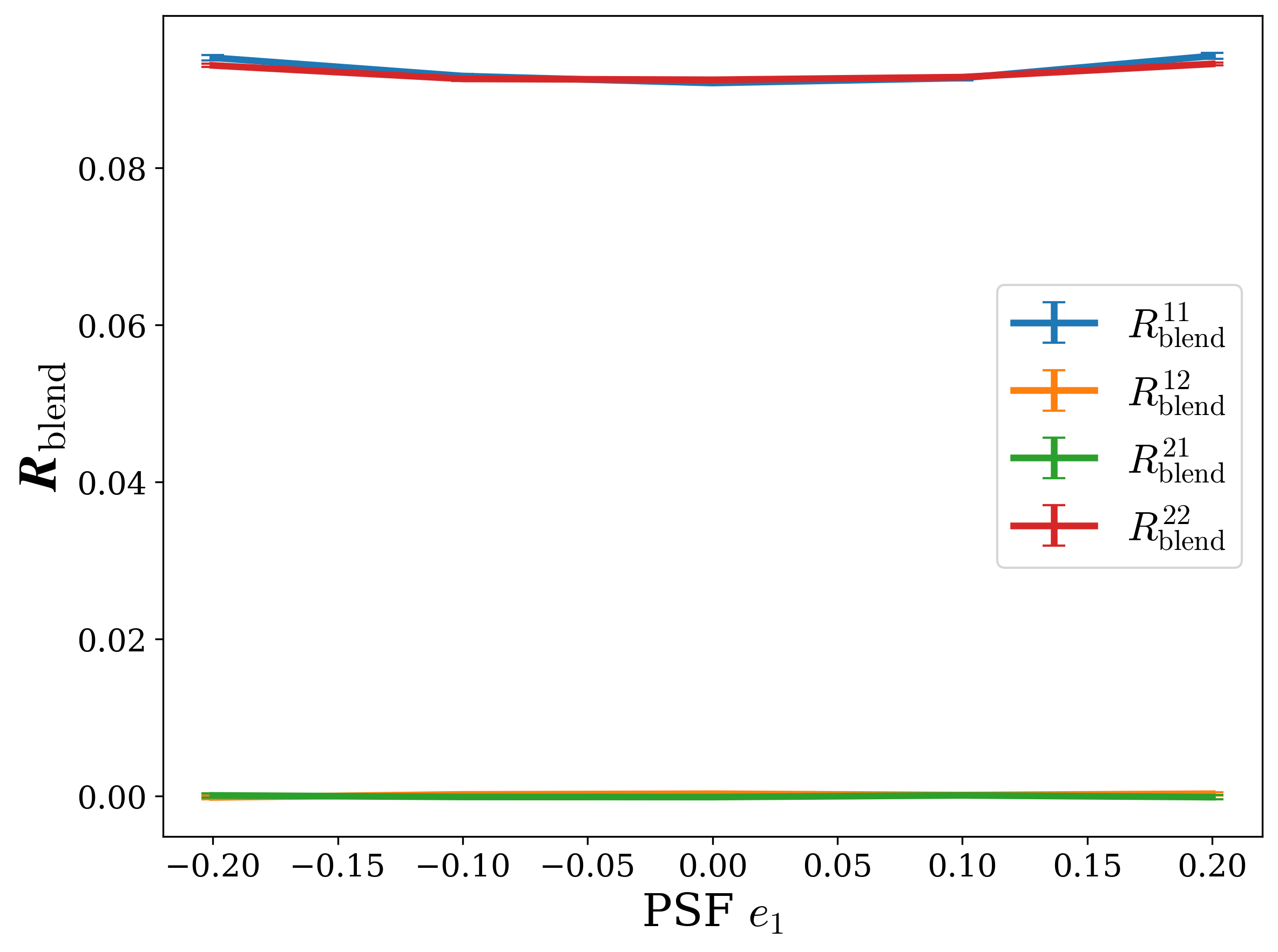}
    \caption{Measurements of $\bm{R}_{\rm blend}$ to the neighbouring galaxy when changing the first component of the PSF ellipticity with the other component remaining zero.}
    \label{fig:psf_test}
\end{figure}

Developing an accurate PSF model at each detection position is essential for producing unbiased shear estimates and should be thoroughly evaluated alongside the shape measurement method. In this work, we have employed a constant PSF across the simulated tile and between different tiles. For simplicity, the input PSF was used for correction during shape measurement. The impact of the error in the reconstructed PSF only goes into the individual shape measurement. However, we acknowledge that the PSF can influence blending in several ways. The size of the PSF affects the level of blending and the de-blending process, potentially shifting the scale at which the aforementioned selection effect of ambiguous blends becomes significant. Regarding galaxy shape measurements, the PSF size and ellipticity also influence the response to neighbouring galaxies. We have partially addressed this issue by scaling the length properties input into our emulators by the factor of the PSF $r_e$. Nonetheless, it remains necessary to examine how PSF ellipticity impacts $R_{\rm blend}$.

Similar to Appendix \ref{sec:linearity}, we conduct tests using stamp-based toy simulations. Using the same configuration of a galaxy pair with an identical primary galaxy and the first secondary galaxy, we sequentially varied one component of the PSF ellipticity while keeping the other component fixed at zero. As in the linearity test, $\bm{R}_{\rm blend}$ was measured by changing the shear applied to the secondary galaxy for each ellipticity value. The secondary galaxy is rotated around the primary from $0$ to $2\pi$ in steps of $\pi/2$. The measured shapes are averaged over 10 realisations and rotation angles. The results are shown in Fig. \ref{fig:psf_test}. All response components remain unaffected by variations in the first component of PSF ellipticity. Results for the second component are not shown, as they exhibit similar behaviour. A 10\% change in $R_{\rm blend}$ is observed for a 30\% change in PSF ellipticity, though such levels of PSF distortion are rare. While the exact amplitude depends on the specific configuration of the blend, this test suggests that the blending shear response is relatively insensitive to PSF anisotropy and its variations.

\section{Generating a mock catalogue}
\label{sec:catalogue_app}

In this section, we describe in detail how we created our galaxy catalogue. We first drew a catalogue of galaxies' physical properties from observationally informed priors using the \textsc{Prospector}-{$\beta$} \citep{wangWangInferringMoreLess2023} model. Those properties include total stellar mass, redshift and star formation rates in logarithmically adjacent cosmic time bins. Their priors were established with a motivation to put joint and correlated constraints on redshift and galaxy properties, thus breaking the degeneracies between the SED template colours and redshift. Apart from the physically informed priors, Prospector-beta samples additional parameters from flat priors that are needed to model the dust attenuation, the gas and the AGN emission. A more detailed description of the priors can be found in \cite{wangWangInferringMoreLess2023} and in Sect. 2 of \cite{Luca2024Stellar}. 

We have used the same fiducial samples as in \cite{Luca2024Stellar}, where $10^6$ sets of galaxy properties are drawn from the \textsc{Prospector}-{$\beta$} model. In their work, galaxy SEDs are generated for each object by providing the sampled properties as input to the stellar population synthesis code \textsc{FSPS} \citep{Conroy2009,Conroy2010}. The SEDs are then integrated in the KiDS-VIKING filter bands to get the corresponding AB magnitudes. To minimise the computation cost of generating new SEDs, we expand the base catalogue to the desired size by sampling more physical sources from \textsc{Prospector}-{$\beta$} priors and assigning them the existing photometry properties of the closest neighbour from the base catalogue based on all the mentioned physical properties. To avoid duplicate values that could potentially cause overfitting in the training of the emulators, we add Gaussian stochasticities to the photometry parameters with \cite{griffithAdvancedCameraSurveys2012} measurement variances.

The morphology model of galaxies is also required to simulate galaxy images. Parametric models like single or mixed Sersic profiles are the most typical options, with more sophisticated alternatives such as observed galaxy images and generative models being increasingly common. We use a single Sérsic profile in this work, although we acknowledge that more realistic models may be of interest. Utilising an inference algorithm named vine copulas and COSMOS observations, \cite{liKiDSLegacyCalibrationUnifying2023} developed a learned model that captures the mutual distributions among galaxy morphological parameters, magnitudes and redshifts. {Using their model, we generate the structural properties of the samples, including $n_{\rm sersic}$, $R_{\rm e}$, and axis ratio, based on each galaxy’s physical properties. In particular, stellar mass and other \textsc{Prospector}-{$\beta$} parameters are sampled first and propagated forward to determine magnitudes, which the morphology model conditions on.}

Finally, the catalogue is resampled conditioned on galaxy stellar mass to match a target stellar mass function (SMF). The latter is the one presented in Prospector beta and \textsc{Prospector}-{$\alpha$} \cite{lejaNewCensus022020}, which has been obtained by fitting the 3D-HST and COSMOS2015 observations. The SMF parametrisation in \cite{lejaNewCensus022020} is the sum of two redshift-dependent Schechter functions \citep{lejaNewCensus022020}. This guarantees a smoothly evolving SMF with cosmic time. We take the median of the fitting results from \cite{lejaNewCensus022020} as our fiducial SMF model throughout the paper. Essentially, an accurate determination of the $n_{\gamma}$ correction depends heavily on the right assumption of the true source galaxy population. Therefore, it is necessary to examine the impact of the uncertainties in SMF in terms of how they change the photometry properties of galaxies used to determine $R$, which we will explore in Sect. \ref{r_model}.

\section{Boosted trees}
\label{sec:trees_math}

Tree-based ML models are among the most popular algorithms in data mining. Starting from the single-tree models, for example, a decision tree sequentially splits the training data at each node into two subdivisions based on one feature. The tree develops its nodes and branches by optimising the objective function. Variance reduction is one of the most commonly deployed methods for regression problems, which measures the purity increase of the data in the two new nodes compared to the old node. Similarly, in classification problems, cross-entropy (also known as LogLoss) is used as the metric to measure the impurity of the nodes. A node stops splitting as a leaf at given hyperparameters such as maximum tree depth, minimum data points in each leaf and so on. The leaf value calculates the average of the training data that reside in it and returns it as the prediction for new data that falls into the leaf. This implies that the predictions are deterministic and discrete.

Random forests average the outputs of numerous independent decision trees trained with different random seeds and optionally different subsets of the features, making it possible to generate continuous predictions. Similarly, instead of parallelising the trees, gradient boosting builds them in a successive way where each tree learns the error residuals of previous trees to achieve an ultimate smaller bias. Rather than averaging, the prediction of gradient boosting is an additive function of every tree's outputs. Another shared component by the two models is column subsampling (or feature subsampling), which uses only a random subset of the data features to train each tree.

Here we follow the exposition as in \cite{chenXGBoostScalableTree2016}. The optimisation of the boosted trees searches for the best structure of each tree successively using a greedy algorithm that grows the branches starting from the root leaf based on the gain after splitting. A new tree keeps being added in this way until the model loss converges. In its second-order approximation, the regularised objective function optimising the current tree $f_t$ is written as,
\begin{equation}
\label{eq:loss}
    \mathcal{L}^{(t)} \simeq \sum_i [l(y_i, \hat{y}^{(t-1)}_i) + g_i~f_t(\textbf{x}_i) + \frac{1}{2} h_i~f_t^2(\textbf{x}_i)] + {\rm \Omega}(f_t),
\end{equation}
where $i$ iterates over the training samples. $\hat{y}^{(t-1)}_i$ is the prediction of the model composed of the $(t-1)$ previous trees while $y_i$ is the sample value and $\textbf{x}_i$ is the corresponding sample features. In the second term, $\Omega(f_t)$ is the regularisation that penalises the complexity of the current tree and mitigates over-fitting, often taking the form of
\begin{equation}
    \Omega = \gamma T + \frac{1}{2} \lambda \sum_{j=1}^T \omega^2_j,
\end{equation}
where $\gamma$ and $\lambda$ are hyperparameters determined before optimisation, while $T$ is the number of leaves with the score of each denoted as $\omega_j$. The optimal values for $\omega_j$ given a fixed tree structure can be found by minimising the regularised loss of the model as Eq. \ref{eq:loss}, and then we end up having the corresponding optimal scoring $\it{L}^{(t)}$ for the structure,
\begin{equation}
    \mathcal{L}^{(t)} = -\frac{1}{2} \sum_{j=1}^T S_j + \gamma T
\end{equation}
and
\begin{equation}
    S_j = \frac{(\sum_{i \in I_j} g_i)^2}{\sum_{i \in I_j} h_j + \lambda},
\end{equation}
where $\sum_{i \in I_j}$ sums up the samples that go into leaf $j$. Finally, similar to the aforementioned impurity decreasing technique seen in the regular tree models, the tree is built up from the root by sequentially splitting the leaf $k$ into leaf $k+1$ and leaf $k+2$, achieving the highest gain
\begin{equation}
    {\rm Gain} = \frac{1}{2} (S_{k+1}+S_{k+2}-S_{k}) - \gamma.
\end{equation}

\section{Emulator invariance under varying observing conditions}
\label{sec:invariance}

\begin{figure*}
    \centering
    \includegraphics[width=0.9\linewidth]{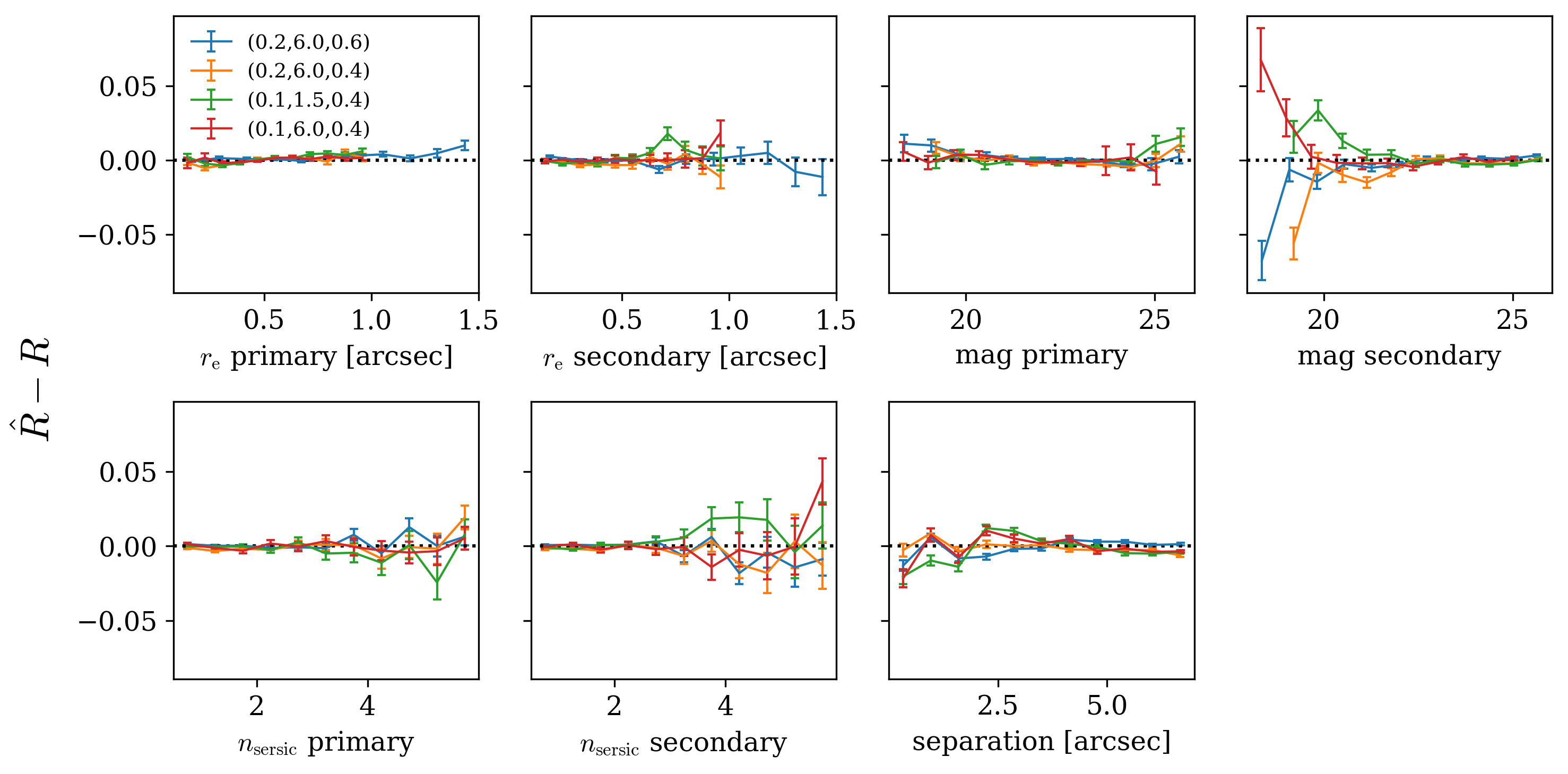}
    \caption{Residuals between model-predicted $R_{\rm blend}$ and the measurements across four sets of simulations with varying observing conditions. Numbers in brackets in the legend are pixel scale, pixel noise, and PSF FWHM of each set, in the units of $\mathrm{arcsec}$, $\mathrm{counts}$ per pixel, and $\mathrm{arcsec}$, respectively.}
    \label{fig:R_vary}
\end{figure*}

\begin{table}
  \centering
  \caption{Averaged shear response from model predictions $\hat{R}$ against measurements $R$ across four sets of simulations with varying observing conditions. The model is always trained on (0.2,6.0,0.6). The balanced Brier score losses (BBSL) of the detection emulator are shown in the last row. A BBSL of 0 indicates perfect classification, whereas uniformly random predictions in the range (0, 1) would result in BBSL $\approx 0.333$.}
  \label{tab:R_vary}
      \begin{tabular}{l S[table-format=1.4(3)] S[table-format=1.4(3)] S[table-format=1.3]}
      \toprule
      & \multicolumn{1}{c}{$R_{\rm blend}\times10^2$} & \multicolumn{1}{c}{$\hat{R}_{\rm blend}\times10^2$} & BBSL \\
      \midrule
      {(0.2,6.0,0.6)} & \num{1.0787(375)} & \num{1.1401(14)} & 0.062 \\
      {(0.2,6.0,0.4)} & \num{0.9722(486)} & \num{0.7870(18)} & 0.105 \\
      {(0.1,1.5,0.4)} & \num{0.8990(536)} & \num{0.8386(18)} & 0.088 \\
      {(0.1,6.0,0.6)} & \num{0.5064(652)} & \num{0.4944(23)} & 0.061 \\
      \bottomrule
    \end{tabular}
    
\end{table}

We simulated four simulations with varying relevant observing conditions, including pixel scale, background noise level, and PSF FWHM. They are (0.2,6.0,0.6), (0.2,6.0,0.4), (0.1,1.5,0.4) and (0.1,6.0,0.6), in the unit of $\mathrm{arcsec}$, $\mathrm{counts}$ per pixel, and $\mathrm{arcsec}$, respectively, with 200 tiles for the first set and 100 tiles for the rest. We compare the measurements and predictions from the model trained on (0.2,6.0,0.6).

We find that pairs with very close and very bright neighbours introduce a $10$–$20\%$ bias in the overall $\bar{R}_{\rm blend}$ during these variation tests, likely due to the non-linearity of the response in this regime. Excluding detections that have another detection within 3 arcseconds that is at least 5 times brighter effectively mitigates this bias, and results in only a $\sim2\%$ reduction in the sample. In practice, such a filter can be applied during the catalogue selection stage. When evaluating the redshift correction from the input catalogue, where we have a detection probability $p_i$ for each input galaxy $i$ and $p_{j,|3~{\rm arcsec}}$ for its neighouring galaxies $j$ within 3 arcsec, its probability of being detected and not to be removed by this bright neighbour is then given as
\begin{equation}
    p_{\rm stay,i} = p_i \prod_j (1-p_{j,|3~{\rm arcsec}}),
\end{equation}
where the product runs over the neighbouring galaxies that are 5 times brighter than the primary galaxy. $\Sigma p_{\rm stay,i}$ over all input galaxies removes roughly $3\%$ detections compared to $\Sigma p_i$. 

The residuals of the shear response between model predictions and measurements from varying simulations are shown in Fig. \ref{fig:R_vary}. The mean responses of galaxy pairs are shown in Table \ref{tab:R_vary}. No significant bias is observed for observation settings that are different from the training data. This is despite the fact that the dynamical range of the mean response is an order of magnitude greater than the uncertainty of its measurement in a simulation. We note that the scaled parameter limits can exceed the ones in the training data, in which case pairs are removed to avoid extrapolation. This leads to a loss of pairs, although most of them lie in the regions where $R$ vanishes for large pair separations or the detection probability becomes negligible due to their faintness. A fraction of small sources was also rejected, whereas their impact is small due to the fraction and the fact that they are poorly resolved in the first place. The balanced Brier score losses of the detection model are also presented across the sets in Table \ref{tab:R_vary}. The model exhibits a slightly weaker yet reasonably good performance on simulations with settings that differ from the training data.


\label{lastpage}
\end{document}